\documentclass[twocolumn,showpacs,preprintnumbers,amsmath,amssymb,aps,prl]{revtex4}

\usepackage{graphicx}
\usepackage{dcolumn}
\usepackage{bm}
\usepackage{amsmath,amssymb}
\usepackage{epstopdf}

\begin{document}

\preprint{}

\title{Broadband Phonon Scattering in PbTe-based Materials Driven Near the Ferroelectric Phase Transition by Strain or Alloying}
\author{Ronan M. Murphy\textsuperscript{1,2}}
\author{\'Eamonn D. Murray\textsuperscript{2}}
\author{Stephen Fahy\textsuperscript{1,2}}
\author{Ivana Savi\'c\textsuperscript{2}}
\email{ivana.savic@tyndall.ie}
\affiliation{\textsuperscript{\normalfont{1}}Department of Physics, University College Cork, College Road, Cork, Ireland}
\affiliation{\textsuperscript{\normalfont{2}}Tyndall National Institute, Dyke Parade, Cork, Ireland}

\date{\today}

\begin{abstract}

The major obstacle in the design of materials with low lattice thermal conductivity is the difficulty in efficiently scattering phonons across the entire frequency spectrum. Using first principles calculations, we show that driving PbTe materials to the brink of the ferroelectric phase transition could be a powerful strategy to solve this problem. We illustrate this concept by applying tensile [001] strain to PbTe and its alloys with another rock-salt IV-VI material, PbSe; and by alloying PbTe with a rhombohedral IV-VI material, GeTe. This induces extremely soft optical modes at the zone center, which increase anharmonic acoustic-optical coupling and decrease phonon lifetimes at all frequencies. We predict that PbTe, Pb(Se,Te) and (Pb,Ge)Te alloys driven close to the phase transition in the described manner will have considerably lower lattice thermal conductivity than that of PbTe (by a factor of $2\--3$). The proposed concept may open new opportunities for the development of more efficient thermoelectric materials. 

\end{abstract}


\maketitle

\section{Introduction} 

The efficient control of thermoelectric energy conversion processes is highly desirable as nearly $60$\% of all energy consumed worldwide is wasted in the form of heat~\cite{energyflow}. Low thermal conductivity, together with high electrical conductivity and Seebeck coefficient, leads to a high thermoelectric figure of merit $ZT$ of a material~\cite{nnano8-471,advmat22-3970,kanatzidis-chemmat,nmat7-105}. In semiconducting materials, heat propagates mostly via phonons with mean free paths ranging from the nm to $\mu$m scale~\cite{nmat7-105,kappa,massdisorder}. This poses a serious issue for design efforts to suppress heat conduction over all relevant length scales. It is particularly difficult to prevent the propagation of low-frequency phonons with long mean free paths that carry a significant amount of heat~\cite{kappa,massdisorder}.

The challenge of engineering efficient phonon scattering across the entire spectrum was tackled recently by fabricating complex PbTe and related IV-VI materials with multi-scale hierarchical design~\cite{nat414-418,allscale-pbse,allscale-pbte-mgte,allscale-snte,ees7-251}. The combined action of mesoscale grain boundaries that scatter low-frequency phonons, and nanoscale precipitates and atomic disorder that affect mid- and high-frequencies, decreased the lattice thermal conductivity $\kappa$ of PbTe by a factor of $2\--3$, and raised its $ZT$ above the target value of $2$~\cite{nat414-418,allscale-pbte-mgte,ees7-251}. Optimization of an all-scales design could result in further $\kappa$ reductions, since nanostructured PbTe without mesoscale grain boundaries can exhibit $4-5$ times lower $\kappa$ than that of PbTe~\cite{kanatzidis-chemmat,ultralowkappa1,ultralowkappa2}. However, the electrical conductivity and Seebeck coefficient may be negatively affected in that process due to enhanced electron scattering at the interfaces~\cite{nat414-418,allscale-pbte-mgte,ultralowkappa1,ultralowkappa2}. It is thus desirable to establish alternative concepts to realize efficient phonon scattering in PbTe and related materials throughout the spectrum without degrading their electronic thermoelectric properties.

In this paper, we propose a strategy that could enable effective phonon scattering in the whole spectrum of PbTe materials, and which does not rely on the concept of nanostructuring. It is well known that PbTe crystallizes in the rock-salt structure, and it is energetically close to the ferroelectric phase transition to the rhombohedral structure, which corresponds to the frozen-in atomic motion of the transverse optical (TO) mode along [111] direction~\cite{prb32-2302}. Consequently, TO phonons have relatively low frequencies at the zone center ($\approx 1$ THz), and interact strongly with acoustic modes~\cite{ssc148-417,nmat10-614,prb85-155203}. These effects lead to unusually small lifetimes of TO and acoustic modes, and exceptionally low $\kappa$~\cite{prb85-155203}. These properties of PbTe give rise to the idea that if the proximity to the transition and the anharmonic acoustic-TO interaction could be enhanced, the phonon lifetimes and the $\kappa$ values may become considerably lower than those of PbTe. Although the sensitivity to the volume changes of the calculated TO modes and the $\kappa$ of PbTe has been reported~\cite{ssc148-417,prb80-024304,prb91-214310,prb89-205203}, the thermal transport properties of PbTe materials near the ferroelectric phase transition, and specific proposals how to achieve them, have never been investigated.

Here we show from first principles that efficient scattering for all phonon frequencies can be achieved by driving PbTe and its alloys to the verge of the ferroelectric phase transition. This can be induced by applying tensile [001] strain to PbTe or its alloys with another rock-salt IV-VI material, such as PbSe. We also present an alternative route to drive PbTe materials near the ferroelectric transition without using strain: alloying of PbTe with a rhombohedral IV-VI material, such as GeTe. Enhanced phonon scattering across the spectrum in all these materials arises due to the TO softening associated with the increased proximity to the phase transition, which causes much stronger anharmonic acoustic-TO interaction than in PbTe and Pb(Se,Te) alloys. In addition to this key effect, alloy disorder further scatters high-frequency phonons in (Pb,Ge)Te and strained Pb(Se,Te) alloys near the phase transition, similarly as in Pb(Se,Te) alloys~\cite{pbsete}. As a result, the lattice thermal conductivity of PbTe, Pb(Se,Te) and (Pb,Ge)Te alloys tuned to the close proximity of the phase transition as described will be significantly decreased compared to PbTe (by a factor of $2\--3$). Conductivity reductions of this magnitude would be comparable to those measured in all-scale structured materials with record $ZT$ values~\cite{nat414-418,allscale-pbte-mgte}.

The proposed concept is general, and it would also be applicable to other materials close to soft zone center optical mode transitions. It can be achieved by using strain, alloying, and possibly also by nanostructuring. Our proposal may be realized experimentally, for example, by metastable growth of (Pb,Ge)Te solid solutions~\cite{pbgete}, or by depositing a thin film of PbTe or Pb(Se,Te) alloys on a flexible polymer substrate~\cite{pbs-polymer,pbse-pbs-polymer}, and applying tensile [001] strain directly to the polymer. 

\section{Methodology}

We calculate the lattice thermal conductivity without any adjustable parameters, using density functional theory (DFT)~\cite{rmp64-1045} combined with the Boltzmann transport equation in the relaxation time approximation (BTE-RTA)~\cite{srivastava}. This approach and its variations have been successful in reproducing measured $\kappa$ for a number of materials so far~\cite{apl91-231922,kappa,massdisorder,prb85-155203,prb87-165201}. It can thus be considered as a good predictive tool in designing new materials with tailored thermal properties. In our implementation, harmonic and third order anharmonic interatomic force constants (IFCs) at $0$~K were calculated from Hellman-Feynman forces using a real-space finite difference supercell approach~\cite{kappa} and the Phono3py code~\cite{phonopy,prb84-094302}. All DFT calculations were carried out using the \textsc{abinit} code~\cite{abinit}, and employing the local density approximation (LDA) and Hartwigsen-Goedecker-Hutter norm-conserving pseudopotentials~\cite{hgh}. Forces were computed on $216$ atom supercells, using an energy cut-off of $15$ Ha and $4$ shifted $2\times 2\times 2$ reciprocal space grids for electronic states. More calculation details are given in the Supplemental Material~\cite{suppl}. 

Within the BTE-RTA framework, the lattice thermal conductivity (for simplicity, thermal conductivity from here on) is given by~\cite{srivastava}:  
\begin{equation}
\label{kappa}
\kappa=\frac{1}{N_{\bf q}V}\sum_{{\bf q},s}c_{{\bf q},s}v_{{\bf q},s}^2\tau_{{\bf q},s}, 
\end{equation}
where ({\bf q},s) denotes a phonon mode with reciprocal space vector {\bf q} and branch index $s$, $\omega_{{\bf q},s}$ is its frequency, $c_{{\bf q},s}$ the heat capacity, $v_{{\bf q},s}=\text{d}\omega_{{\bf q},s}/\text{d}{\bf q}$ the group velocity, and $\tau_{{\bf q},s}$ the lifetime. $V$ is the primitive cell volume and $N_{\bf q}$ is the ${\bf q}$ point grid size (we used $40\times 40\times 40$ grids). We calculated anharmonic phonon lifetimes using Fermi's Golden Rule and taking into account the contribution of three-phonon scattering processes~\cite{srivastava}. Details of our BTE-RTA implementation are described in Refs.~\cite{ivana-apl,pccp}. We also verified the applicability of this approach in describing the $\kappa$ of PbTe (see Supplemental Material~\cite{suppl2}). We treated Pb(Se,Te) and (Pb,Ge)Te alloys using the virtual crystal approximation, and accounted for mass disorder perturbatively, via an effective lifetime~\cite{massdisorder,pbsete}, which was then combined with the three-phonon lifetimes using Mattheisen's rule. Treating disorder in the force constants on the same footing as mass disorder in the BTE-RTA $\kappa$ calculations is a challenging problem~\cite{pccp,pbsete-better-disorder}, and it was not pursued in this work. Therefore, the ratios of $\kappa$ between PbTe and alloyed materials obtained here should be interpreted as the lower limit to their actual values~\cite{klemens-point-def}.

\section{PbTe materials driven near the ferroelectric phase transition} 

\begin{figure}
  \begin{center}
    \includegraphics[width=0.45\textwidth]{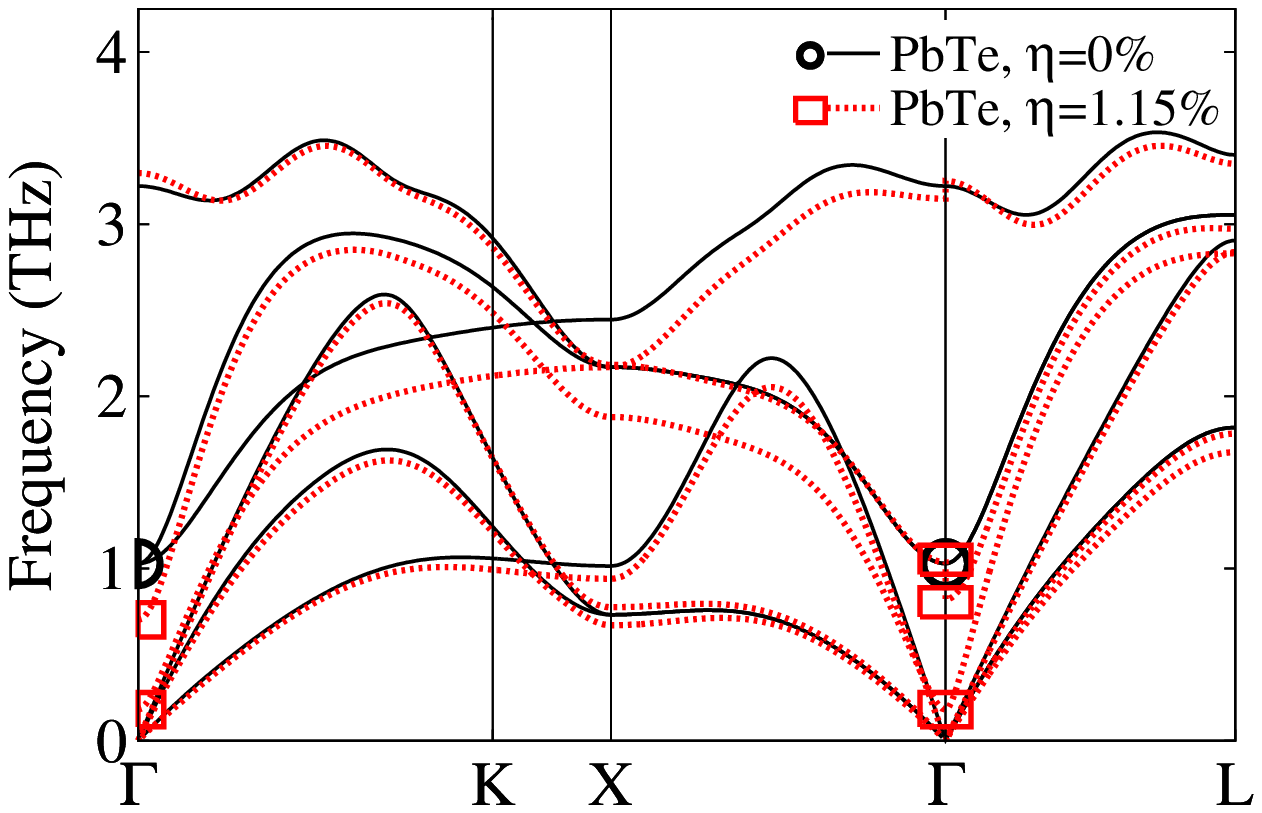}
    \includegraphics[width=0.45\textwidth]{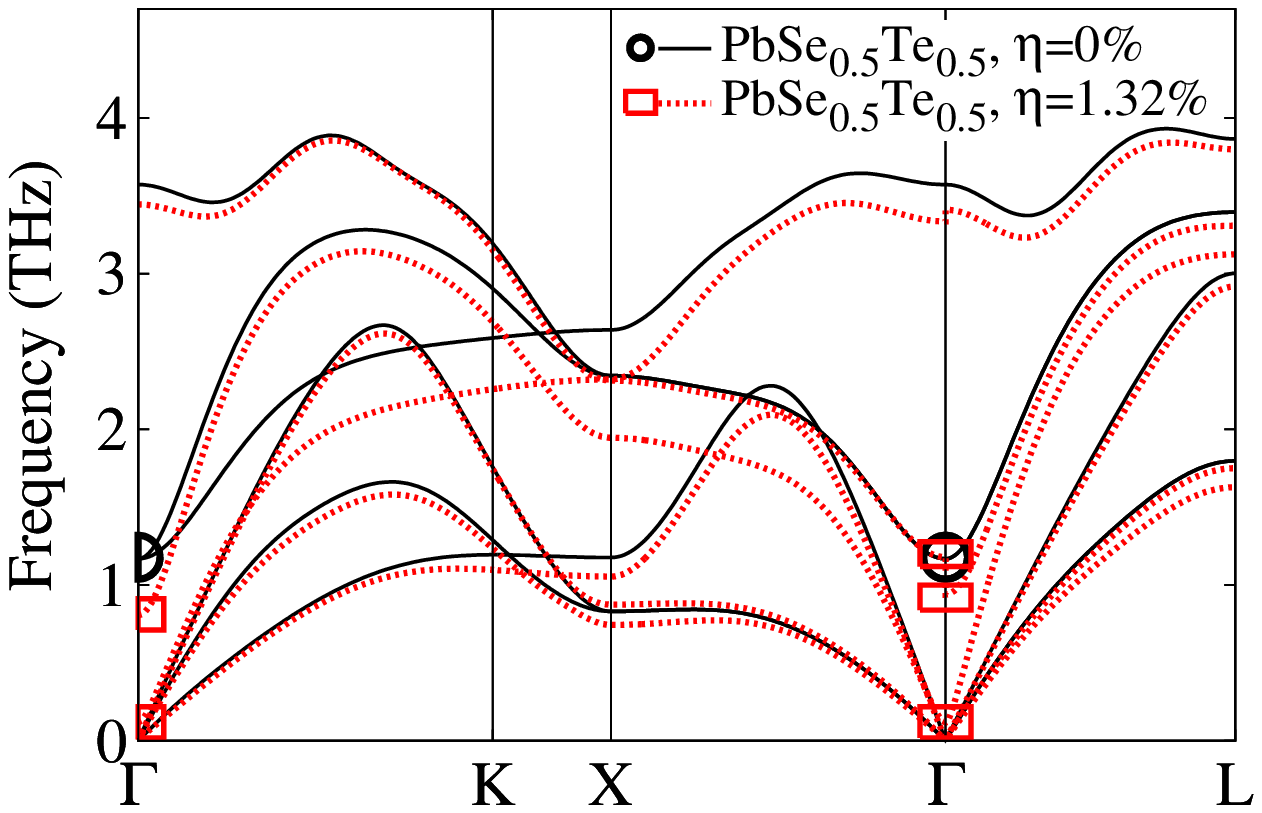}
    \includegraphics[width=0.45\textwidth]{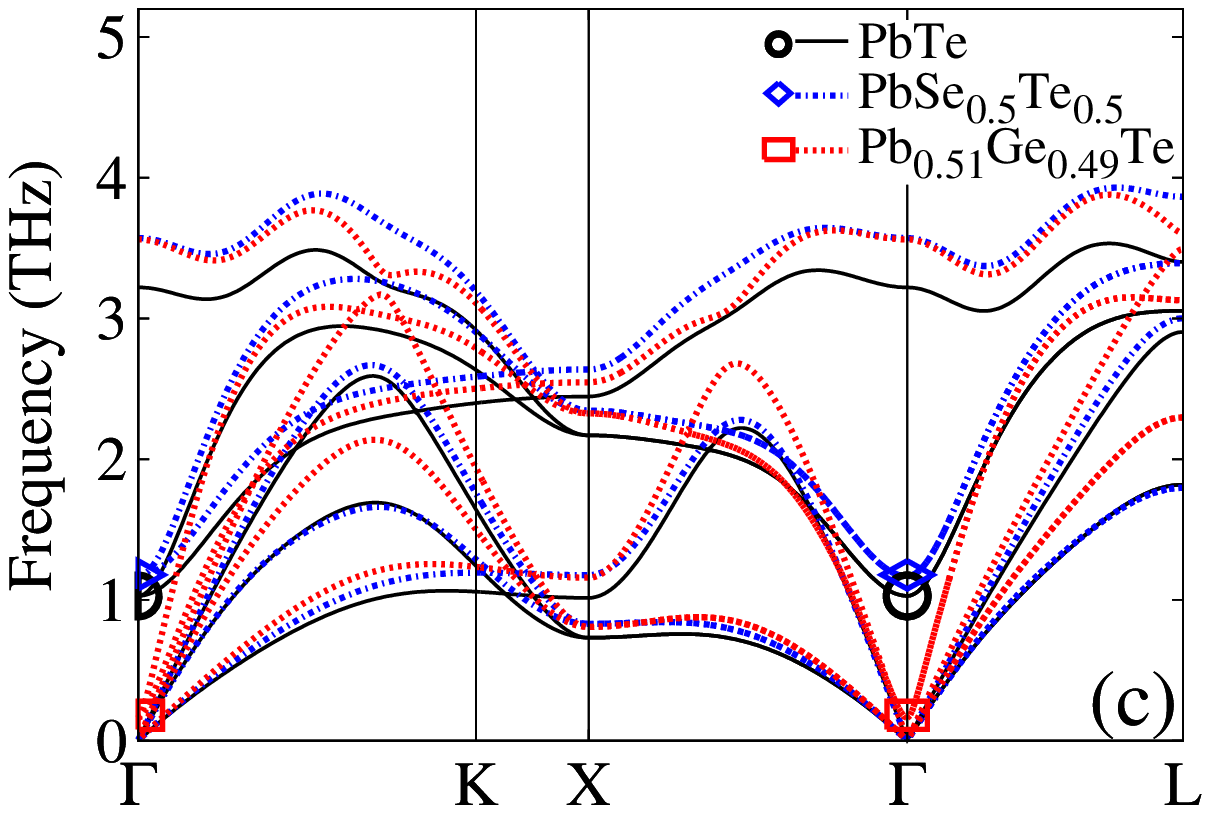}
  \end{center}
  \caption{(color online) (a) Phonon dispersions at $0$~K for PbTe (solid black lines) and PbTe driven to the verge of the phase transition by tensile [001] strain of $\eta=1.15$\% (dotted red lines). Frequencies of the soft transverse optical modes at the zone center, TO($\Gamma$), are represented with circles for PbTe, and rectangles for PbTe driven to the transition. (b)~Phonon dispersions at $0$~K for PbSe$_{0.5}$Te$_{0.5}$ (solid black lines) and PbSe$_{0.5}$Te$_{0.5}$ driven to the verge of the phase transition by tensile [001] strain of $\eta=1.32$\% (dotted red lines).  TO($\Gamma$) frequencies are shown in circles for PbSe$_{0.5}$Te$_{0.5}$, and rectangles for PbSe$_{0.5}$Te$_{0.5}$ driven to the transition. (c) Phonon dispersions at $0$~K for PbTe (solid black lines), PbSe$_{0.5}$Te$_{0.5}$ (dash-dotted blue lines), and Pb$_{0.51}$Ge$_{0.49}$Te alloy at the verge of the phase transition (dotted red lines). TO($\Gamma$) frequencies are represented with circles for PbTe, diamonds for  PbSe$_{0.5}$Te$_{0.5}$, and rectangles for Pb$_{0.51}$Ge$_{0.49}$Te. Phonon dispersions for alloys were calculated using the virtual crystal approximation.}
  \label{fig1}
\end{figure}

{\it Strained PbTe.} To determine the amount of strain which will push PbTe to the verge of the phase transition, we calculated the TO phonon frequencies at the zone center as a function of [001] strain. Strain was simulated by constraining the lattice constant $a_{\parallel}$ in the $[100]$ and $[010]$ direction, and relaxing the lattice constant $a_{\perp}$ in the $[001]$ direction. The amount of strain is defined as $\eta=(a_{\parallel}-a_{0})/a_{0}$, where $a_0$ is the equilibrium lattice constant of PbTe obtained from DFT relaxation. Since $[001]$ strain reduces the symmetry of the rock-salt structure to tetragonal, the degeneracy of the two TO modes is lifted. (Strictly speaking, these modes do not have pure TO character, but for simplicity, we will use this term in the rest of the paper). We found that $\eta=1.15$\% will soften one of the TO modes from the equilibrium value of $\sim 1$~THz down to $\sim 0.1$~THz, and consequently bring PbTe close to the phase transition. This is illustrated in Fig.~\ref{fig1}~(a) that compares the phonon band structures of PbTe (solid black lines) and strained PbTe at the verge of the phase transition (dotted red lines) along high symmetry lines for the cubic symmetry. The other TO mode will also become softer due to applied strain, but to a lesser extent. The frequencies of the optical modes in strained PbTe close to the zone center show directional dependence, which is a consequence of the non-analytic nature of the ion-ion interaction~\cite{prb50-13035}.

{\it Strained Pb(Se,Te) alloys.} PbSe$_x$Te$_{1-x}$ alloys can be tuned near the phase transition in a similar manner as PbTe, by changing the amount of applied tensile [001] strain. Since these alloys are formed by mixing of two rock-salt materials, their TO frequencies are bounded by those of PbTe and PbSe ($1$ THz and $1.2$ THz, respectively). Subsequently, they cannot be driven to the phase transition by changing the alloy composition, and they need to be strained to achieve this effect. Here we focused on the $x=0.5$ composition since it has the lowest $\kappa$~\cite{pbsete}, but the same strategy can be applied to any other composition. Our calculations show that an extremely soft TO mode can be induced in PbSe$_{0.5}$Te$_{0.5}$ using tensile [001] strain of $\eta=1.32$\%, see Fig.~\ref{fig1}~(b).

{\it (Pb,Ge)Te alloys.} The proximity to the ferroelectric phase transition of PbTe materials can be dramatically increased not only by applying strain to PbTe and its alloys with other rock-salt IV-VI materials, but also by alloying PbTe with other rhombohedral IV-VI materials, e.g. GeTe. (Pb,Ge)Te alloys are markedly different from Pb(Se,Te) alloys because they undergo the ferroelectric phase transition from the rhombohedral to the rock-salt structure with temperature~\cite{pbgete}. Consequently, their proximity to the phase transition and softening of TO modes can be dramatically increased by varying the alloy composition and temperature, unlike in the case of Pb(Se,Te). The temperature at which the transition between the two phases occurs in Pb$_{1-x}$Ge$_x$Te alloys decreases as the Ge content decreases, from $\sim 670$~K at $x=1$ down to $0$~K at $x\approx 0.01$~\cite{pbgete}. 

The temperature that brings Pb$_{1-x}$Ge$_x$Te alloy near the phase transition for any composition $x>0.01$ could be determined by computing the temperature dependence of the TO frequencies at the zone center. Conversely, one could determine the amount of Ge that drives Pb$_{1-x}$Ge$_x$Te close to the transition for any temperature $T<670$~K by calculating the same frequencies as a function of the alloy composition. Our approach, however, cannot capture the temperature effects on phonon frequencies due to the zero temperature representation of structural properties and IFCs. Nevertheless, it should describe well the qualitative changes of the $\kappa$ of Pb$_{1-x}$Ge$_x$Te alloys at the brink of the phase transition with respect to that of PbTe as argued in the following. Our calculations show that varying the alloy composition within our model does induce the ferroelectric phase transition. We find that degenerate TO modes of Pb$_{1-x}$Ge$_x$Te become much softer than those of PbTe and Pb(Se,Te) alloys ($\sim 0.1$~THz) for the alloy composition of $x=0.49$, as shown in Fig.~\ref{fig1}~(c). Since the transition temperature of Pb$_{0.51}$Ge$_{0.49}$Te is $\sim 450$~K~\cite{pbgete}, such phonon band structure and the associated transitions should describe reasonably well the $\kappa$ of Pb$_{0.51}$Ge$_{0.49}$Te alloy close to that temperature.

\section{Impact on phonon lifetimes} 

\begin{figure}
  \begin{center}
    \includegraphics[width=0.45\textwidth]{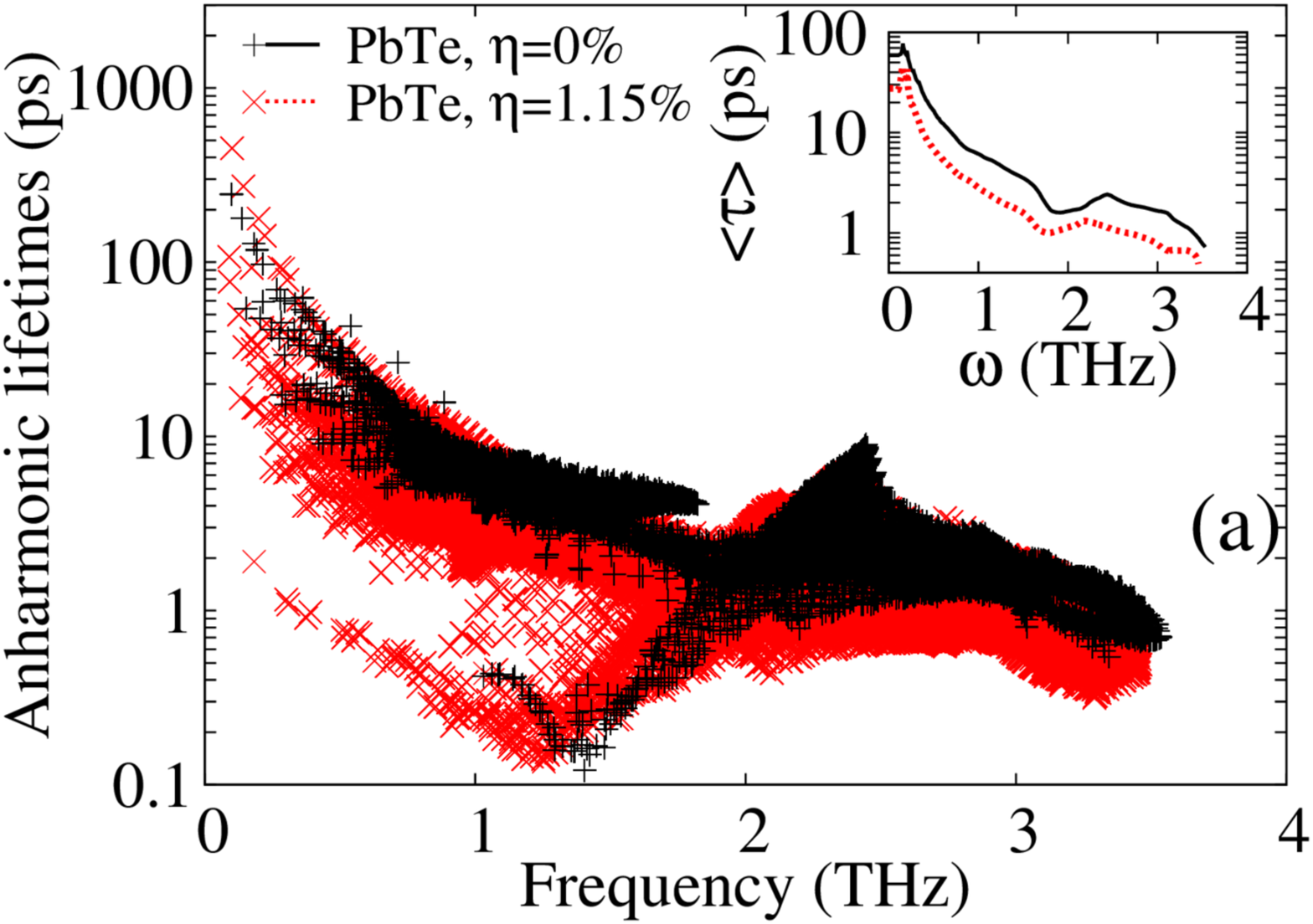}
    \includegraphics[width=0.45\textwidth]{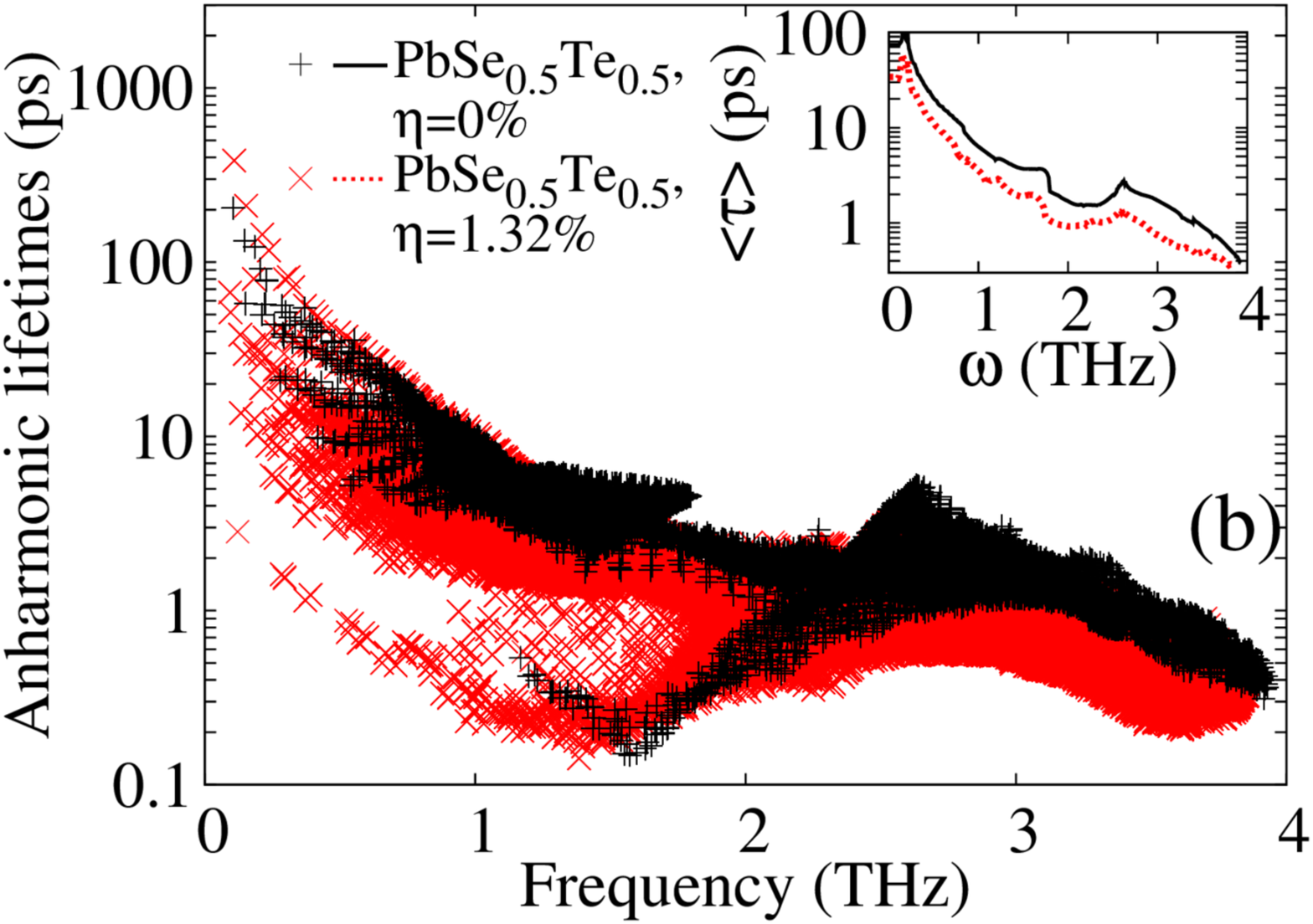}
    \includegraphics[width=0.45\textwidth]{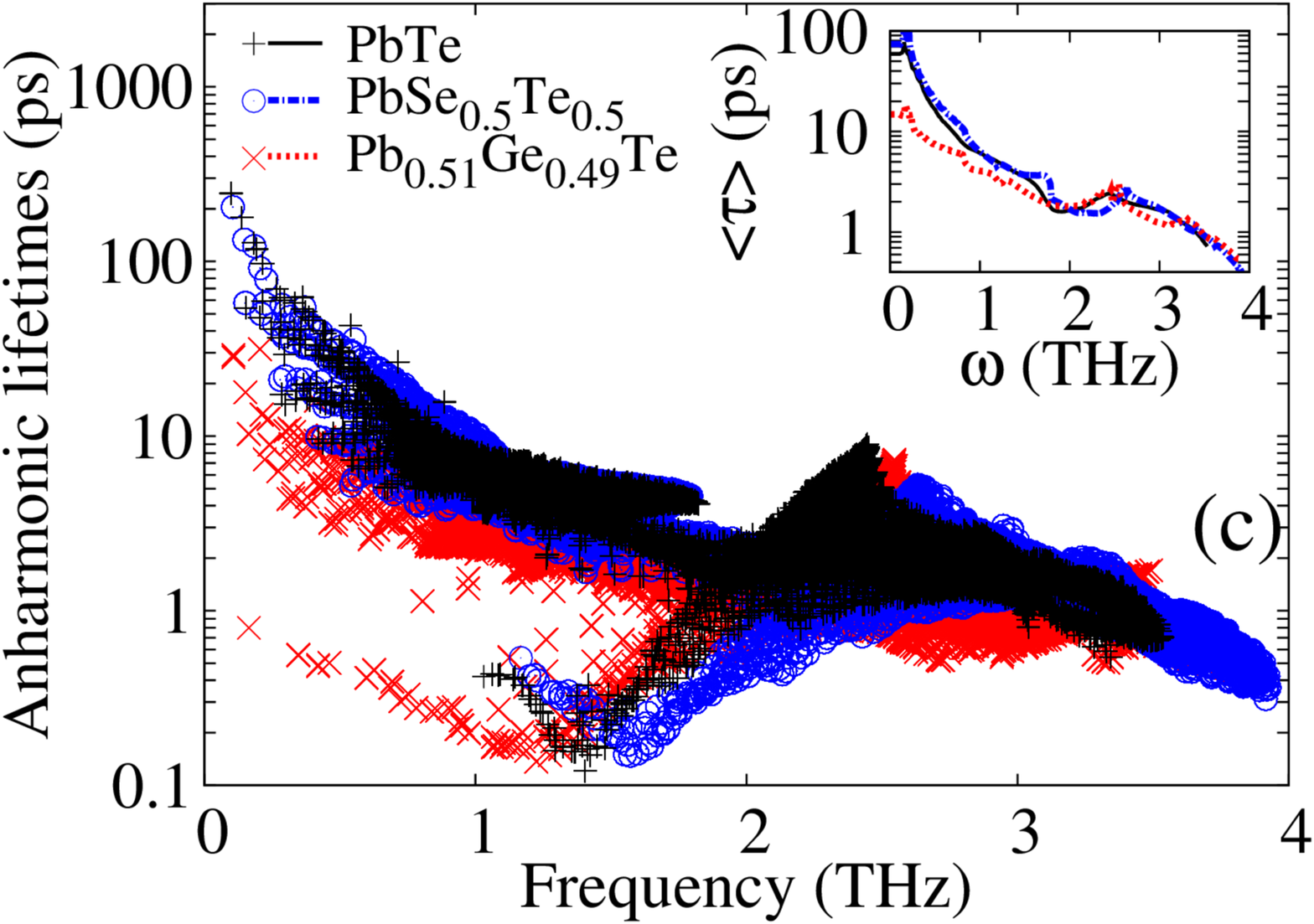}
  \end{center}
  \caption{(color online) (a) Anharmonic (three-phonon) lifetimes at $300$ K as a function of frequency for PbTe (black pluses) and PbTe driven to the verge of the phase transition by tensile [001] strain of $\eta=1.15$\% (red crosses). Inset: Frequency dependence of lifetimes averaged over frequency at $300$ K for PbTe (solid black line) and PbTe driven to the transition (dotted red line). (b) Anharmonic lifetimes at $300$ K and their frequency averages (inset) as a function of frequency for PbSe$_{0.5}$Te$_{0.5}$ (black pluses and solid line) and PbSe$_{0.5}$Te$_{0.5}$ driven to the verge of the phase transition by tensile [001] strain of $\eta=1.32$\% (red crosses and dotted line). (c) Anharmonic lifetimes at $300$ K and their frequency averages (inset) as a function of frequency for PbTe (black pluses and solid line), PbSe$_{0.5}$Te$_{0.5}$ (blue circles and dash-dotted line), and Pb$_{0.51}$Ge$_{0.49}$Te alloy at the verge of the phase transition (red crosses and dotted line).}
  \label{fig2}
\end{figure}

{\it Strained PbTe.} The strain-engineered phonon band structure of PbTe that maximizes the TO($\Gamma$) softening decreases the phonon lifetimes $\tau$ roughly by a factor of $\sim 2$  with respect to PbTe throughout the spectrum. This is shown in Fig.~\ref{fig2}~(a) and its inset, which represent the anharmonic (three-phonon) lifetimes at $300$~K and their averaged values over frequency, respectively. Remarkably, the lifetimes of strained PbTe are smaller than those of PbTe even for low frequency phonons that are difficult to scatter with commonly used strategies to reduce $\kappa$, such as nanostructuring. Recently, it has been argued that low frequency phonons were efficiently scattered in multiple-scales structures by mesoscale grain boundaries~\cite{nat414-418,allscale-pbse,allscale-pbte-mgte,allscale-snte}. Our strategy achieves the same effect without any need for complex hierarchical design.

The large decrease in the phonon lifetimes of PbTe driven to the phase transition via strain is a direct consequence of the softening of TO modes and their increased anharmonic coupling with acoustic modes. To demonstrate this effect, we computed the acoustic-TO contribution to the total linewidth (inverse of the lifetime) for all frequencies by accounting for the triplets of interacting states that contain at least one acoustic and one TO mode. For each wave vector, we labeled the two lowest phonon modes as transverse acoustic (TA) modes, and the highest mode as longitudinal optical (LO) mode. Since the ordering of TO and longitudinal acoustic (LA) modes changes throughout the Brillouin zone in all the materials considered (see Fig.~\ref{fig1}), we distinguished between them using the following procedure. We determined which one of those three states is mostly longitudinal by projecting their eigenvectors onto the corresponding wave vector, and classified it as LA mode, while the other two states were labeled as TO modes. We found that the acoustic-TO contribution to the linewidth dominates over the other contributions in both equilibrium and strained PbTe across the spectrum. It accounts for $\sim 70$\% of the linewidth of acoustic and LO modes (except for the $\sim 2-3$~THz range), and for nearly $100$\% of the linewidth of TO modes (see Supplemental Material~\cite{suppl2}). Furthermore, by straining PbTe near the phase transition, the acoustic-TO contribution to the linewidth typically increases by a factor of $2\--3$ throughout the spectrum, and up by a factor of $10\--100$ for some frequencies (see Supplemental Material~\cite{suppl2}). We conclude that extremely soft TO modes generated by strain considerably increase the anharmonic acoustic-TO interaction, which in turn substantially reduces the phonon lifetimes. 

Both the increased phase space and coupling strength 
are responsible for the increase of the acoustic-TO contribution to the total linewidth by applying strain to PbTe. The phase space for the three-phonon scattering is related to the energy and momentum conservation of these processes~\cite{threephonon}. The coupling strength can be quantified using the expression for the three-phonon linewidth without the energy conservation terms~\cite{kappa}. Our analysis of the phase space and the coupling strength associated with the acoustic-TO interaction similar to that for the linewidths shows that they both increase by straining PbTe, 
and thus lead to the increase of the acoustic-TO contribution to the linewidth. 

We also found that a relatively small number of soft TO modes near the zone center that interact strongly with acoustic modes play a disproportionally large role in determining the total linewidth of PbTe strained near the phase transition. To illustrate this, we calculated the contribution to the anharmonic linewidth due to the coupling of acoustic modes with the TO modes within the sphere centered at $\Gamma$ with the radius of $1/3$ of the $\Gamma$-X distance ($\sim 1/27$ volume of the Brillouin zone), here labeled as TO$_1$ modes. They represent only $\sim 1.3$\% of the total number of modes, and the channels that involve both acoustic and TO$_1$ modes contribute only $\sim 3$\% to the total scattering phase space across the spectrum. In contrast, our calculations show that the anharmonic acoustic-TO$_1$ interaction has a comparatively much larger effect on the linewidth in strained PbTe, as a result of the large coupling strength. The acoustic-TO$_1$ contribution accounts for $\sim 20\--30$\% of the linewidth throughout the spectrum, and for $\sim 100$\% of the linewidth of TO$_1$ modes (see Supplemental Material~\cite{suppl2}). Additionally, 
by driving PbTe near the phase transition via strain,
the acoustic-TO$_1$ contribution to the linewidth typically increases by a factor of $2\--3$ across the spectrum, and up by a factor of $10^2\--10^7$ for some frequencies (see Supplemental Material~\cite{suppl2}). These findings reveal an important contribution of 
ultra soft TO modes near $\Gamma$ in increasing the anharmonic acoustic-TO interaction and reducing the phonon lifetimes in strained PbTe.

{\it Strained PbSe$_{0.5}$Te$_{0.5}$ alloy.} The anharmonic lifetimes of PbSe$_{0.5}$Te$_{0.5}$ alloy driven near the phase transition via strain are also significantly reduced with respect to those of PbSe$_{0.5}$Te$_{0.5}$, see Fig.~\ref{fig2}~(b). A similar analysis as in the case of strained PbTe shows that the softer TO modes and the stronger acoustic-TO/TO$_1$ interaction in strained PbSe$_{0.5}$Te$_{0.5}$ are the main reason for the considerable decrease of its anharmonic lifetimes with respect to PbSe$_{0.5}$Te$_{0.5}$. Strained PbSe$_{0.5}$Te$_{0.5}$ alloys provide an additional advantage for enhanced phonon scattering compared to strained PbTe. Mass disorder scatters high-frequency phonons more efficiently than the three-phonon interaction, similarly as in PbSe$_{0.5}$Te$_{0.5}$~\cite{pbsete}, further reducing the effective lifetimes at high frequencies (see Supplemental Material~\cite{suppl2}).

{\it Pb$_{0.51}$Ge$_{0.49}$Te alloy.} The anharmonic lifetimes of Pb$_{0.51}$Ge$_{0.49}$Te alloy near the phase transition are also considerably lower than those of PbTe and PbSe$_{0.5}$Te$_{0.5}$. The TO frequencies of PbTe and PbSe$_{0.5}$Te$_{0.5}$ are comparable (see Fig.~\ref{fig1}~(c)), which leads to their similar acoustic-TO and three-phonon contributions to $\tau$. The later effect is illustrated in Fig.~\ref{fig2}~(c) that compares the anharmonic lifetimes of PbTe (black pluses) and PbSe$_{0.5}$Te$_{0.5}$ (blue circles).
In sharp contrast, Pb$_{0.51}$Ge$_{0.49}$Te alloy is energetically much closer to the phase transition, which results in its much lower TO frequencies near the zone center (Fig.~\ref{fig1} (c)) and reduced anharmonic $\tau$ across the spectrum and particularly at low frequencies (red crosses in Fig.~\ref{fig2}~(c)). This finding further confirms that extremely soft TO modes near $\Gamma$ have a highly beneficial role in effective scattering of a wide range of phonon frequencies. The analysis of the acoustic-TO/TO$_1$ contributions to the anharmonic linewidth of Pb$_{0.51}$Ge$_{0.49}$Te with respect to PbTe shows that their increase is indeed responsible for the reduced $\tau$ in Pb$_{0.51}$Ge$_{0.49}$Te (see Supplemental Material~\cite{suppl2}), largely due to the increased coupling strength. As in strained PbSe$_{0.5}$Te$_{0.5}$, mass disorder in Pb$_{0.51}$Ge$_{0.49}$Te alloy is more efficient in scattering high-frequency phonons than the phonon-phonon interaction, which results in an additional $\tau$ decrease (see Supplemental Material~\cite{suppl2}).

\section{Impact on thermal conductivity} 

\begin{figure}
  \begin{center}
    \includegraphics[width=0.45\textwidth]{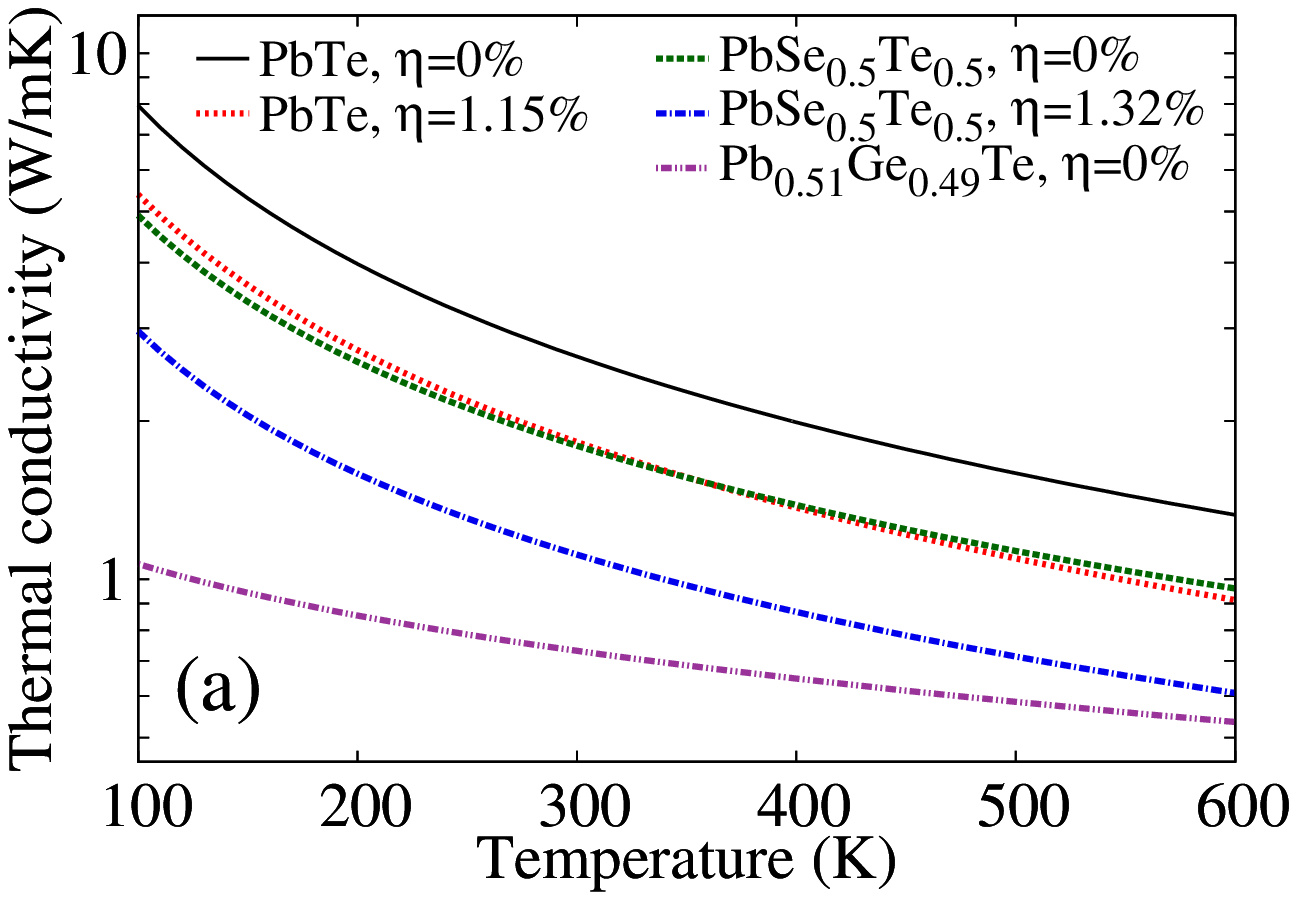}
    \includegraphics[width=0.45\textwidth]{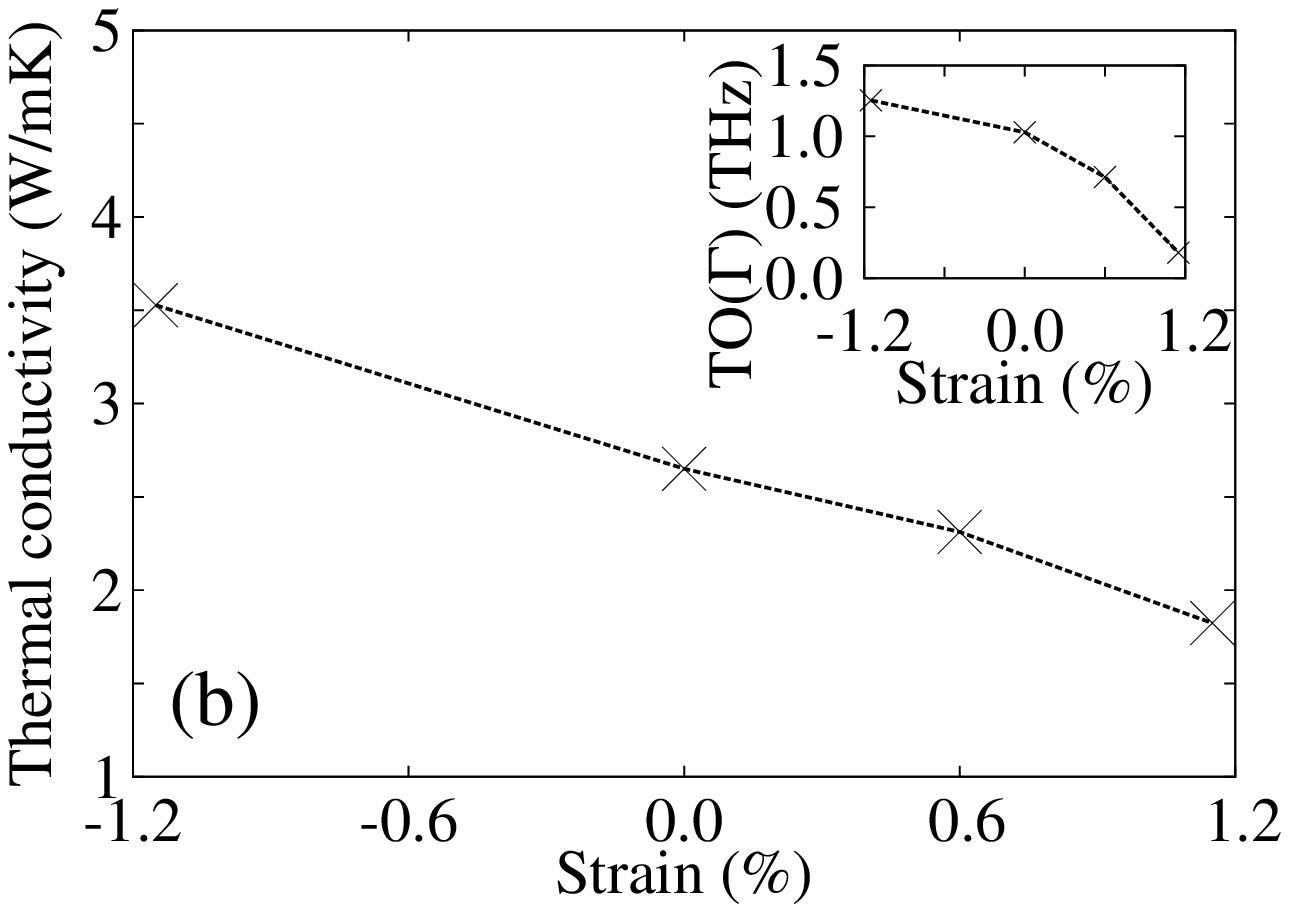}
    \includegraphics[width=0.45\textwidth]{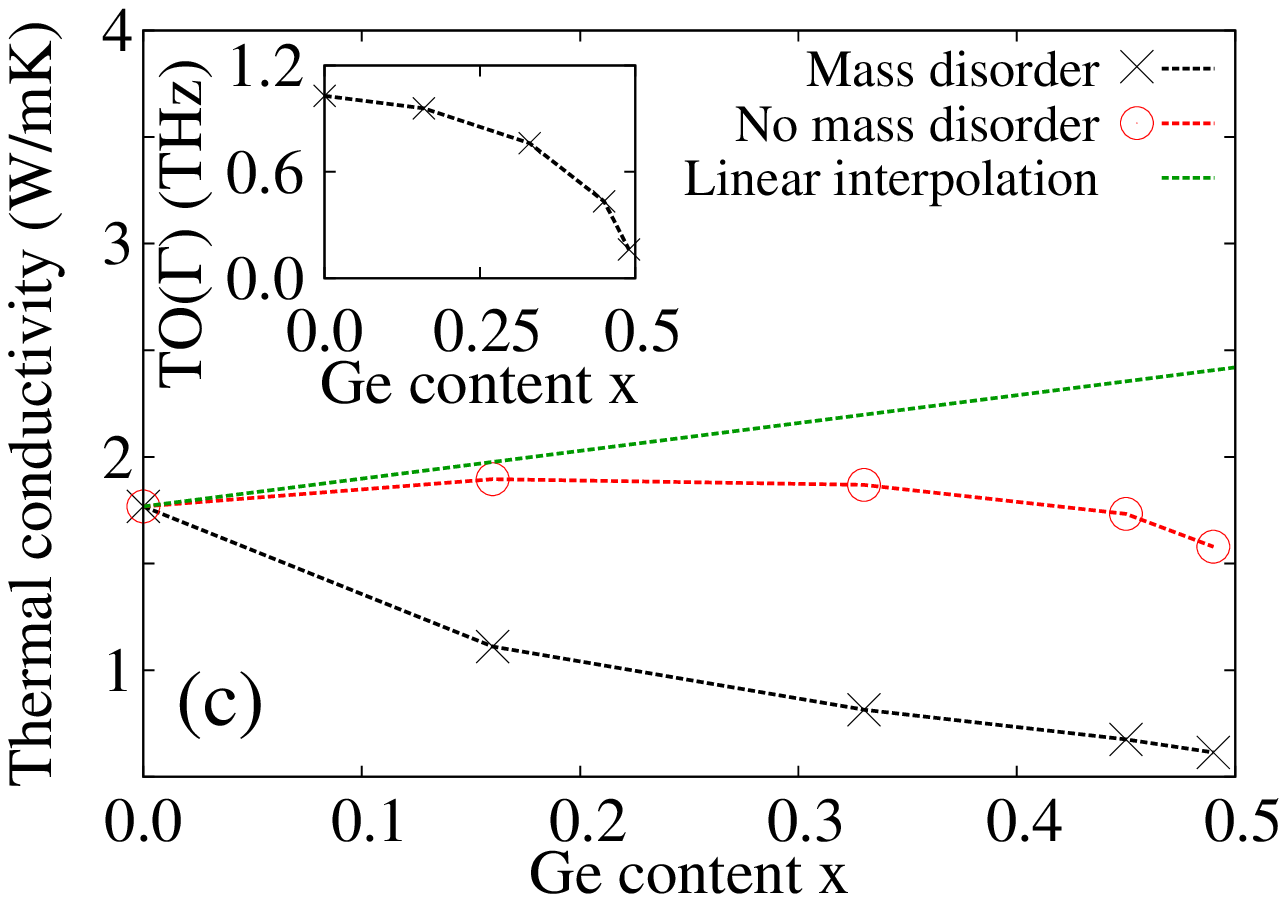}
  \end{center}
  \caption{(color online) (a) Lattice thermal conductivity versus temperature for PbTe (solid black line), PbTe driven near the phase transition by tensile [001] strain of $\eta=1.15$\% (dotted red line), PbSe$_{0.5}$Te$_{0.5}$ (dashed green line), PbSe$_{0.5}$Te$_{0.5}$ driven near the phase transition by tensile [001] strain of $\eta=1.32$\% (dash-dotted blue line), and Pb$_{0.51}$Ge$_{0.49}$Te alloy near the phase transition (dash-double dotted purple line). Conductivity for strained materials in shown along the out-of-plane direction. (b) Out-of-plane lattice thermal conductivity of PbTe at $300$~K as a function of [001] strain. Inset: Lowest transverse optical mode frequency at the zone center, TO($\Gamma$), 
for PbTe at $0$~K versus [001] strain. (c) Lattice thermal conductivity of Pb$_{1-x}$Ge$_x$Te alloys at $450$~K as a function of Ge content $x$: full calculation (black crosses), calculation without mass disorder (red circles) and linearly interpolated values between those calculated for PbTe and GeTe (dashed green line). Inset: TO($\Gamma$) frequency of Pb$_{1-x}$Ge$_x$Te at $0$~K versus Ge content $x$.}
  \label{fig4}
\end{figure}

{\it Strained PbTe.} A factor of $\sim 2$ reduction of the phonon lifetimes at all frequencies in PbTe strained near the phase transition leads to the reduction in the out-of-plane thermal conductivity by a factor of $1.5$ with respect to equilibrium. This is represented in Fig.~\ref{fig4}~(a) for the temperature range $100-600$~K, where solid black and dotted red lines correspond to the $\kappa$ of equilibrium and strain-driven PbTe, respectively. The difference in the $\tau$ and $\kappa$ reductions is due to the larger group velocities of some of the phonon modes in strained PbTe. Interestingly, in the in-plane direction, this difference becomes very small and the $\kappa$ decreases by a factor of $1.9$. We also computed the out-of-plane $\kappa$ dependence on the amount of [001] strain, see Fig.~\ref{fig4}~(b) for $300$~K. The $\kappa$ decrease directly correlates with the TO($\Gamma$) softening shown in the inset, in agreement with our conclusion that softer TO modes lead to more effective phonon scattering. 

The efficiency of the proposed concept in reducing the thermal conductivity of PbTe is comparable to that of alloying with PbSe. Our computed $\kappa$ of PbSe$_{0.5}$Te$_{0.5}$ (dashed green line in Fig.~\ref{fig4}~(a)) is also a factor of $1.5$ lower than that of PbTe, in agreement with a previous calculation~\cite{pbsete}. Accounting for disorder in the force constants would increase the value of this factor~\cite{pccp,pbsete-better-disorder,klemens-point-def} closer to experiments ($\sim 2$ at $300$~K~\cite{pbte-exp,pbsete-exp}). We emphasize that two very different mechanisms cause the low $\kappa$ in these two types of materials. Strained PbTe benefits from the enhanced anharmonic acoustic-TO interaction 
across the spectrum, while alloy disorder in PbSe$_{0.5}$Te$_{0.5}$ blocks the flow of high-frequency phonons only. 

The low thermal conductivity of strained PbTe originates from the softer TO modes near the zone center and their increased anharmonic interaction with acoustic modes with respect to equilibrium. To illustrate this effect more quantitatively, we artificially replaced the acoustic-TO$_1$ contribution to the total linewidth of strained PbTe with that of PbTe in its $\kappa$ calculation at $300$~K. The $\kappa$ values obtained using this procedure are $\sim 7$\% larger than those of strained PbTe. This result shows that TO$_1$ modes, which contribute only $\sim 1.3$\% to the total number of modes but interact strongly with acoustic phonons, produce a much stronger effect on the $\kappa$ than one would expect from their relative number.

{\it Strained PbSe$_{0.5}$Te$_{0.5}$ alloy.} The strategy of exploiting extremely soft TO modes to reduce the thermal conductivity of PbTe will become as effective as all-scale structuring~\cite{nat414-418,allscale-pbte-mgte,ees7-251} if it is combined with alloying. This approach simultaneously incorporates the mechanisms of enhanced anharmonic acoustic-TO interaction and alloy disorder to reduce $\kappa$ more effectively. We find that driving PbSe$_{0.5}$Te$_{0.5}$ to the brink of the phase transition via tensile [001] strain will reduce its out-of-plane $\kappa$ by a factor of at least $2.4$ with respect to PbTe (dash-dotted blue line in Fig.~\ref{fig4}~(a)). If we neglect mass disorder in our calculations, the $\kappa$ of strained PbSe$_{0.5}$Te$_{0.5}$ is $1.5$ times lower than that of PbTe (similarly to strained PbTe) due to the increased acoustic-TO interaction. However, mass disorder is more efficient in scattering high-frequency phonons of strained PbSe$_{0.5}$Te$_{0.5}$, and causes an additional $\kappa$ decrease by a factor of $1.6$ compared to its values without mass disorder. The actual $\kappa$ reduction of strained PbSe$_{0.5}$Te$_{0.5}$ with respect to PbTe will be even larger than the factor of $2.4$ computed here, due to disorder in the force constants~\cite{klemens-point-def}.

{\it Pb$_{0.51}$Ge$_{0.49}$Te alloy.} The concept of combining 
ultra soft TO modes and alloying to substantially reduce the thermal conductivity of PbTe can also be realized without strain e.g. by tuning the composition of Pb$_{1-x}$Ge$_x$Te alloys to bring them near the phase transition at desired temperatures. These conditions can be achieved only in alloys of PbTe with another rhombohedral material, such as GeTe. As already discussed, Pb$_{0.51}$Ge$_{0.49}$Te alloy remains close to the transition at all temperatures in our calculations, thus describing the $\kappa$ appropriately only near the transition temperature of $\sim 450$~K~\cite{pbgete}. Our results show that the $\kappa$ of Pb$_{0.51}$Ge$_{0.49}$Te alloy (dash-double dotted purple line in Fig.~\ref{fig4}~(a)) will be at least $2.9$ times lower than the $\kappa$ of PbTe at $450$~K, and even lower than that of strained PbSe$_{0.5}$Te$_{0.5}$. 

The very low thermal conductivity of Pb$_{0.51}$Ge$_{0.49}$Te alloy stems from both the increased anharmonic acoustic-TO interaction and strong alloy scattering. Fig.~\ref{fig4}~(c) and its inset show the $\kappa$ values and the TO($\Gamma$) frequencies of Pb$_{1-x}$Ge$_x$Te alloys at $450$~K as a function of the alloy composition $x$ for $0\le x \le 0.49$ (black crosses). This figure also shows the corresponding $\kappa$ computed by neglecting mass disorder (red circles). We note that if soft TO modes were not induced in Pb$_{1-x}$Ge$_{x}$Te alloys, their $\kappa$ without mass disorder would roughly be equal to the linearly interpolated values between the $\kappa$'s of their parent materials PbTe and GeTe, as e.g. in the case of Pb(Se,Te).
The interpolated values between our calculated $\kappa$ for PbTe and GeTe at $450$~K are shown in dashed green line in Fig.~\ref{fig4}~(c)~\cite{kappa-gete}. 
The increased TO softening in Pb$_{1-x}$Ge$_{x}$Te with $x$ leads to the larger reduction of the $\kappa$ values that neglect mass disorder with respect to the interpolated values, by up to a factor of $1.6$ for $x=0.49$. Additionally, the relatively large mass difference between Pb and Ge atoms results in a large $\kappa$ decrease with respect to the case when mass disorder is neglected, by up to a factor of $2.6$ for Pb$_{0.51}$Ge$_{0.49}$Te~\cite{kappa-vs-t-pbgete}. We conclude that the extra softening of TO modes in Pb$_{1-x}$Ge$_{x}$Te alloys with respect to PbTe together with strong mass disorder makes them more suitable materials for achieving low $\kappa$ compared to PbSe$_x$Te$_{1-x}$ alloys.

The group velocities of acoustic modes in Pb$_{1-x}$Ge$_x$Te alloys increase considerably as the Ge content increases, unlike the group velocities of PbTe and PbSe$_{0.5}$Te$_{0.5}$ with the increasing amount of tensile [001] strain (see Fig.~\ref{fig1}). This is the reason why the $\kappa$  of Pb$_{1-x}$Ge$_x$Te without mass disorder does not decrease monotonously with $x$, as opposed to the $\kappa$ of increasingly strained PbTe, see Figs.~\ref{fig4}~(b) and (c). Furthermore, the $\kappa$ values without mass disorder for Pb$_{1-x}$Ge$_x$Te are larger than those of strained PbTe and PbSe$_{0.5}$Te$_{0.5}$ because of the larger group velocities.

Similarly to strained PbTe, a small number of soft TO modes close to the zone center, which interact more strongly with acoustic modes as Pb$_{1-x}$Ge$_{x}$Te alloys are driven to the phase transition by increasing $x$, have a relatively large effect on their thermal conductivity. We quantify this by substituting the acoustic-TO$_1$ contribution to the total anharmonic linewidth of Pb$_{0.51}$Ge$_{0.49}$Te with that of equilibrium PbTe in its $\kappa$ calculation that neglects mass disorder. This results in the $\sim 18$\% larger value at $300$~K than that of Pb$_{0.51}$Ge$_{0.49}$Te, which is even larger than the corresponding increase in strained PbTe.

\section{Discussion} 

We analyzed the limiting cases where PbTe based materials with the rock-salt and tetragonal structures are driven close to the ferroelectric phase transition, with the goal of evaluating the largest thermal conductivity reductions that could be obtained using this strategy. Since the transition from the rock-salt to the rhombohedral structure is nearly the second order phase transition~\cite{PhysRevB.36.6631}, further increases of the Ge content in Pb$_{1-x}$Ge$_x$Te above $x=0.49$ should only increase the thermal conductivity. 
We also note that the mechanical stability of the tensile strained materials very near the transition may be compromised. Nevertheless, the proposed concept will hold even if the materials are not strained very close to the transition, although their $\kappa$ reductions will be smaller, as illustrated in Fig.~\ref{fig4}~(b).

Even if higher order anharmonicity terms were non-negligible for PbTe materials driven to the transition, the proposed concept would remain valid. However, the $\kappa$ reductions would be lower than our results, particularly at higher temperatures. The main implication of higher anharmonic terms on the thermal properties of PbTe is shifting the TO modes upwards (resulting in the well known TO frequency increase with temperature)~\cite{nmat10-614,prb91-214310,bozin,marianetti,PhysRevLett.112.175501}. This will weaken the anharmonic acoustic-TO interaction, which may result in somewhat higher $\kappa$ values than those predicted here. This effect may become more evident for $T>300$~K in the case of PbTe~\cite{prb91-214310}, and it may manifest at even lower $T$ for the materials near the phase transition. Consequently, the actual $\kappa$ reduction may decrease with $T$ with respect to our calculations.

Experimentally, Pb$_{0.51}$Ge$_{0.49}$Te 
undergoes the phase transition from the rhombohedral to the rock-salt structure at $\sim 450$~K~\cite{pbgete}, while it is rock-salt for all temperatures in our calculations. Higher order anharmonicity terms would further stabilize the rock-salt structure in our virtual crystal simulations~\cite{prb91-214310}, which indicates that using an alloy model beyond this approximation may be necessary to accurately describe the temperature behavior of the $\kappa$ of Pb$_{0.51}$Ge$_{0.49}$Te. This, however, will not change the main implication of our work that tuning the composition of Pb$_{1-x}$Ge$_x$Te alloys to drive the material near the phase transition at a certain temperature will substantially decrease its $\kappa$ at that temperature. The measured $\kappa$ values of some of the (Pb,Ge)Te materials reported in Ref.~\cite{jac621-345} indeed exhibit pronounced dips near the phase transition temperature, and thus support our argument.

Low thermal conductivity and high figure of merit have been recently reported in (Pb,Ge)Te materials with phase separated regions and nanoscale features~\cite{pbgete-jacs,gelbstein-adv-energy-mat,solidstatesci}. Since no dips in the $\kappa$ values close to the transition temperature have been observed in these materials, their $\kappa$ reduction is likely dominated by phonon scattering at the interfaces. Our study shows that substantial $\kappa$ reductions can also be obtained in random (Pb,Ge)Te alloys without any phase separation, which may also be beneficial for electronic transport properties. More systematic studies across all compositions could resolve whether random alloys would be more efficient thermoelectric materials than nanostructured/phase separated materials for certain composition ranges.

\section{Electronic properties}

Driving PbTe materials near the ferroelectric phase transition via strain could preserve their high electrical conductivity and Seebeck coefficient while suppressing the thermal conductivity. Our DFT calculations show that tensile [001] strain of $\eta\sim 1$\% does not change appreciably the electronic band structure close to the conduction/valence band minimum/maximum. Symmetry analysis~\cite{PhysRevB.12.650} and our density functional perturbation theory~\cite{dfpt-abinit} calculations show that the electron-phonon matrix elements for the conduction band maximum or the valence band minimum with TO modes at $\Gamma$ are zero. This indicates that the electronic transport properties of PbTe strained close to the phase transition may not be deteriorated by the TO softening. 
 
Driving (Pb,Ge)Te and strained Pb(Se,Te) alloys close to the phase transition could also be beneficial for their electronic transport properties and thermoelectric figure of merit. It has been recently demonstrated that tuning the alloy composition of PbSe$_x$Te$_{1-x}$ alloys to align the two topmost valence band maxima at a desired temperature increases their electronic conductivity and Seebeck coefficient~\cite{pbsete-nature}. This effect combined with the low $\kappa$ of PbSe$_x$Te$_{1-x}$ alloys leads to significantly increased $ZT$~\cite{pbsete-nature}. Such ``valence band convergence'' mechanism also appears to be responsible for high $ZT$ of Pb$_{0.13}$Ge$_{0.87}$Te nanostructured materials~\cite{pbgete-jacs}. These observations suggest that the band convergence concept may also be realized in both Pb(Ge,Te) and strained Pb(Se,Te) alloys driven to the brink of the phase transition, and result in high $ZT$ values. The proposed strategy to reduce the thermal conductivity of PbTe materials could also be combined with other advanced concepts that improve their electronic transport properties and increase the figure of merit e.g. creating distortions of the electronic density of states close to the Fermi energy via resonant impurity levels~\cite{resimp}.

\section{Summary} 

We predict that driving PbTe materials to the verge of the ferroelectric phase transition via tensile [001] strain or alloying will be an effective strategy to impede phonon flow in the entire spectrum, and thus to considerably reduce the lattice thermal conductivity. The proposed concept is based on the induced softening of the transverse optical modes at the zone center, whose increased anharmonic interaction with acoustic modes enhances phonon scattering. Our first principles calculations show that the efficiency of this approach in reducing the lattice thermal conductivity of Pb(Se,Te) and (Pb,Ge)Te alloys could be comparable to that of all-scale hierarchical architecturing~\cite{nat414-418}. The presented strategy is general, and it would also be applicable to other materials close to soft zone center optical mode transitions. Indeed, a very recent study~\cite{delaire-snse} argued that a similar effect is responsible for the low thermal conductivity of another high-performing thermoelectric material, SnSe~\cite{nature-snse}.

\section{Acknowledgement}

This work was supported by Science Foundation Ireland and the Marie-Curie Action COFUND under Starting Investigator Research Grant 11/SIRG/E2113. \'E. D. M. acknowledges support from EU Commission under Marie Curie International Incoming Fellowship No.~329695. S. F. acknowledges support by Science Foundation Ireland under Grant No. 12/1A/1601. We acknowledge the Irish Centre for High-End Computing (ICHEC) for the provision of computational facilities. We thank Kafil M. Razeeb, David Reis, Mason Jiang and Mariano Trigo for useful discussions.

\bibliography{pbte_strain}

\begin{thebibliography}{66}
\expandafter\ifx\csname natexlab\endcsname\relax\def\natexlab#1{#1}\fi
\expandafter\ifx\csname bibnamefont\endcsname\relax
  \def\bibnamefont#1{#1}\fi
\expandafter\ifx\csname bibfnamefont\endcsname\relax
  \def\bibfnamefont#1{#1}\fi
\expandafter\ifx\csname citenamefont\endcsname\relax
  \def\citenamefont#1{#1}\fi
\expandafter\ifx\csname url\endcsname\relax
  \def\url#1{\texttt{#1}}\fi
\expandafter\ifx\csname urlprefix\endcsname\relax\def\urlprefix{URL }\fi
\providecommand{\bibinfo}[2]{#2}
\providecommand{\eprint}[2][]{\url{#2}}

\bibitem[{ene()}]{energyflow}
\bibinfo{howpublished}{https://flowcharts.llnl.gov/}.

\bibitem[{\citenamefont{Heremans et~al.}(2013)\citenamefont{Heremans,
  Dresselhaus, Bell, and Morelli}}]{nnano8-471}
\bibinfo{author}{\bibfnamefont{J.~P.} \bibnamefont{Heremans}},
  \bibinfo{author}{\bibfnamefont{M.~S.} \bibnamefont{Dresselhaus}},
  \bibinfo{author}{\bibfnamefont{L.~E.} \bibnamefont{Bell}}, \bibnamefont{and}
  \bibinfo{author}{\bibfnamefont{D.~T.} \bibnamefont{Morelli}},
  \bibinfo{journal}{Nat. Nanotechnol.} \textbf{\bibinfo{volume}{8}},
  \bibinfo{pages}{471} (\bibinfo{year}{2013}).

\bibitem[{\citenamefont{Vineis et~al.}(2010)\citenamefont{Vineis, Shakouri,
  Majumdar, and Kanatzidis}}]{advmat22-3970}
\bibinfo{author}{\bibfnamefont{C.~J.} \bibnamefont{Vineis}},
  \bibinfo{author}{\bibfnamefont{A.}~\bibnamefont{Shakouri}},
  \bibinfo{author}{\bibfnamefont{A.}~\bibnamefont{Majumdar}}, \bibnamefont{and}
  \bibinfo{author}{\bibfnamefont{M.~G.} \bibnamefont{Kanatzidis}},
  \bibinfo{journal}{Adv. Mater.} \textbf{\bibinfo{volume}{22}},
  \bibinfo{pages}{3970} (\bibinfo{year}{2010}).

\bibitem[{\citenamefont{Kanatzidis}(2010)}]{kanatzidis-chemmat}
\bibinfo{author}{\bibfnamefont{M.~G.} \bibnamefont{Kanatzidis}},
  \bibinfo{journal}{Chem. Mater.} \textbf{\bibinfo{volume}{22}},
  \bibinfo{pages}{648} (\bibinfo{year}{2010}).

\bibitem[{\citenamefont{Snyder and Toberer}(2008)}]{nmat7-105}
\bibinfo{author}{\bibfnamefont{G.~J.} \bibnamefont{Snyder}} \bibnamefont{and}
  \bibinfo{author}{\bibfnamefont{E.~S.} \bibnamefont{Toberer}},
  \bibinfo{journal}{Nat. Mater.} \textbf{\bibinfo{volume}{7}},
  \bibinfo{pages}{105} (\bibinfo{year}{2008}).

\bibitem[{\citenamefont{Esfarjani et~al.}(2011)\citenamefont{Esfarjani, Chen,
  and Stokes}}]{kappa}
\bibinfo{author}{\bibfnamefont{K.}~\bibnamefont{Esfarjani}},
  \bibinfo{author}{\bibfnamefont{G.}~\bibnamefont{Chen}}, \bibnamefont{and}
  \bibinfo{author}{\bibfnamefont{H.~T.} \bibnamefont{Stokes}},
  \bibinfo{journal}{Phys. Rev. B} \textbf{\bibinfo{volume}{84}},
  \bibinfo{pages}{085204} (\bibinfo{year}{2011}).

\bibitem[{\citenamefont{Garg et~al.}(2011)\citenamefont{Garg, Bonini, Kozinsky,
  and Marzari}}]{massdisorder}
\bibinfo{author}{\bibfnamefont{J.}~\bibnamefont{Garg}},
  \bibinfo{author}{\bibfnamefont{N.}~\bibnamefont{Bonini}},
  \bibinfo{author}{\bibfnamefont{B.}~\bibnamefont{Kozinsky}}, \bibnamefont{and}
  \bibinfo{author}{\bibfnamefont{N.}~\bibnamefont{Marzari}},
  \bibinfo{journal}{Phys. Rev. Lett.} \textbf{\bibinfo{volume}{106}},
  \bibinfo{pages}{045901} (\bibinfo{year}{2011}).

\bibitem[{\citenamefont{Biswas et~al.}(2012)\citenamefont{Biswas, He, Blum, Wu,
  Hogan, Seidman, Dravid, and Kanatzidis}}]{nat414-418}
\bibinfo{author}{\bibfnamefont{K.}~\bibnamefont{Biswas}},
  \bibinfo{author}{\bibfnamefont{J.}~\bibnamefont{He}},
  \bibinfo{author}{\bibfnamefont{I.~D.} \bibnamefont{Blum}},
  \bibinfo{author}{\bibfnamefont{C.-I.} \bibnamefont{Wu}},
  \bibinfo{author}{\bibfnamefont{T.~P.} \bibnamefont{Hogan}},
  \bibinfo{author}{\bibfnamefont{D.~N.} \bibnamefont{Seidman}},
  \bibinfo{author}{\bibfnamefont{V.~P.} \bibnamefont{Dravid}},
  \bibnamefont{and} \bibinfo{author}{\bibfnamefont{M.~G.}
  \bibnamefont{Kanatzidis}}, \bibinfo{journal}{Nature}
  \textbf{\bibinfo{volume}{489}}, \bibinfo{pages}{414} (\bibinfo{year}{2012}).

\bibitem[{\citenamefont{Lee et~al.}(2013)\citenamefont{Lee, Lo, Androulakis,
  Wu, Zhao, Chung, Hogan, Dravid, and Kanatzidis}}]{allscale-pbse}
\bibinfo{author}{\bibfnamefont{Y.}~\bibnamefont{Lee}},
  \bibinfo{author}{\bibfnamefont{S.-H.} \bibnamefont{Lo}},
  \bibinfo{author}{\bibfnamefont{J.}~\bibnamefont{Androulakis}},
  \bibinfo{author}{\bibfnamefont{C.-I.} \bibnamefont{Wu}},
  \bibinfo{author}{\bibfnamefont{L.-D.} \bibnamefont{Zhao}},
  \bibinfo{author}{\bibfnamefont{D.-Y.} \bibnamefont{Chung}},
  \bibinfo{author}{\bibfnamefont{T.~P.} \bibnamefont{Hogan}},
  \bibinfo{author}{\bibfnamefont{V.~P.} \bibnamefont{Dravid}},
  \bibnamefont{and} \bibinfo{author}{\bibfnamefont{M.~G.}
  \bibnamefont{Kanatzidis}}, \bibinfo{journal}{J. Am. Chem. Soc.}
  \textbf{\bibinfo{volume}{135}}, \bibinfo{pages}{5152} (\bibinfo{year}{2013}).

\bibitem[{\citenamefont{Zhao et~al.}(2013)\citenamefont{Zhao, Wu, Hao, Wu,
  Zhou, Biswas, He, Hogan, Uher, Wolverton et~al.}}]{allscale-pbte-mgte}
\bibinfo{author}{\bibfnamefont{L.~D.} \bibnamefont{Zhao}},
  \bibinfo{author}{\bibfnamefont{H.~J.} \bibnamefont{Wu}},
  \bibinfo{author}{\bibfnamefont{S.~Q.} \bibnamefont{Hao}},
  \bibinfo{author}{\bibfnamefont{C.~I.} \bibnamefont{Wu}},
  \bibinfo{author}{\bibfnamefont{X.~Y.} \bibnamefont{Zhou}},
  \bibinfo{author}{\bibfnamefont{K.}~\bibnamefont{Biswas}},
  \bibinfo{author}{\bibfnamefont{J.~Q.} \bibnamefont{He}},
  \bibinfo{author}{\bibfnamefont{T.~P.} \bibnamefont{Hogan}},
  \bibinfo{author}{\bibfnamefont{C.}~\bibnamefont{Uher}},
  \bibinfo{author}{\bibfnamefont{C.}~\bibnamefont{Wolverton}},
  \bibnamefont{et~al.}, \bibinfo{journal}{Energy Environ. Sci.}
  \textbf{\bibinfo{volume}{6}}, \bibinfo{pages}{3346} (\bibinfo{year}{2013}).

\bibitem[{\citenamefont{Tan et~al.}(2014)\citenamefont{Tan, Zhao, Shi, Doak,
  Lo, Sun, Wolverton, Dravid, Uher, and Kanatzidis}}]{allscale-snte}
\bibinfo{author}{\bibfnamefont{G.}~\bibnamefont{Tan}},
  \bibinfo{author}{\bibfnamefont{L.-D.} \bibnamefont{Zhao}},
  \bibinfo{author}{\bibfnamefont{F.}~\bibnamefont{Shi}},
  \bibinfo{author}{\bibfnamefont{J.~W.} \bibnamefont{Doak}},
  \bibinfo{author}{\bibfnamefont{S.-H.} \bibnamefont{Lo}},
  \bibinfo{author}{\bibfnamefont{H.}~\bibnamefont{Sun}},
  \bibinfo{author}{\bibfnamefont{C.}~\bibnamefont{Wolverton}},
  \bibinfo{author}{\bibfnamefont{V.~P.} \bibnamefont{Dravid}},
  \bibinfo{author}{\bibfnamefont{C.}~\bibnamefont{Uher}}, \bibnamefont{and}
  \bibinfo{author}{\bibfnamefont{M.~G.} \bibnamefont{Kanatzidis}},
  \bibinfo{journal}{J. Am. Chem. Soc.} \textbf{\bibinfo{volume}{136}},
  \bibinfo{pages}{7006} (\bibinfo{year}{2014}).

\bibitem[{\citenamefont{Zhao et~al.}(2014{\natexlab{a}})\citenamefont{Zhao,
  Dravid, and Kanatzidis}}]{ees7-251}
\bibinfo{author}{\bibfnamefont{L.-D.} \bibnamefont{Zhao}},
  \bibinfo{author}{\bibfnamefont{V.~P.} \bibnamefont{Dravid}},
  \bibnamefont{and} \bibinfo{author}{\bibfnamefont{M.~G.}
  \bibnamefont{Kanatzidis}}, \bibinfo{journal}{Energy Environ. Sci.}
  \textbf{\bibinfo{volume}{7}}, \bibinfo{pages}{251}
  (\bibinfo{year}{2014}{\natexlab{a}}).

\bibitem[{\citenamefont{Poudeu et~al.}(2006)\citenamefont{Poudeu, D'Angelo,
  Kong, Downey, Short, Pcionek, Hogan, Uher, and Kanatzidis}}]{ultralowkappa1}
\bibinfo{author}{\bibfnamefont{P.~F.~P.} \bibnamefont{Poudeu}},
  \bibinfo{author}{\bibfnamefont{J.}~\bibnamefont{D'Angelo}},
  \bibinfo{author}{\bibfnamefont{H.}~\bibnamefont{Kong}},
  \bibinfo{author}{\bibfnamefont{A.}~\bibnamefont{Downey}},
  \bibinfo{author}{\bibfnamefont{J.~L.} \bibnamefont{Short}},
  \bibinfo{author}{\bibfnamefont{R.}~\bibnamefont{Pcionek}},
  \bibinfo{author}{\bibfnamefont{T.~P.} \bibnamefont{Hogan}},
  \bibinfo{author}{\bibfnamefont{C.}~\bibnamefont{Uher}}, \bibnamefont{and}
  \bibinfo{author}{\bibfnamefont{M.~G.} \bibnamefont{Kanatzidis}},
  \bibinfo{journal}{J. Am. Chem. Soc.} \textbf{\bibinfo{volume}{128}},
  \bibinfo{pages}{14347} (\bibinfo{year}{2006}).

\bibitem[{\citenamefont{Androulakis et~al.}(2007)\citenamefont{Androulakis,
  Lin, Kong, Uher, Wu, Hogan, Cook, Caillat, Paraskevopoulos, and
  Kanatzidis}}]{ultralowkappa2}
\bibinfo{author}{\bibfnamefont{J.}~\bibnamefont{Androulakis}},
  \bibinfo{author}{\bibfnamefont{C.-H.} \bibnamefont{Lin}},
  \bibinfo{author}{\bibfnamefont{H.-J.} \bibnamefont{Kong}},
  \bibinfo{author}{\bibfnamefont{C.}~\bibnamefont{Uher}},
  \bibinfo{author}{\bibfnamefont{C.-I.} \bibnamefont{Wu}},
  \bibinfo{author}{\bibfnamefont{T.}~\bibnamefont{Hogan}},
  \bibinfo{author}{\bibfnamefont{B.~A.} \bibnamefont{Cook}},
  \bibinfo{author}{\bibfnamefont{T.}~\bibnamefont{Caillat}},
  \bibinfo{author}{\bibfnamefont{K.~M.} \bibnamefont{Paraskevopoulos}},
  \bibnamefont{and} \bibinfo{author}{\bibfnamefont{M.~G.}
  \bibnamefont{Kanatzidis}}, \bibinfo{journal}{J. Am. Chem. Soc.}
  \textbf{\bibinfo{volume}{129}}, \bibinfo{pages}{9780} (\bibinfo{year}{2007}).

\bibitem[{\citenamefont{Rabe and Joannopoulos}(1985)}]{prb32-2302}
\bibinfo{author}{\bibfnamefont{K.~M.} \bibnamefont{Rabe}} \bibnamefont{and}
  \bibinfo{author}{\bibfnamefont{J.~D.} \bibnamefont{Joannopoulos}},
  \bibinfo{journal}{Phys. Rev. B} \textbf{\bibinfo{volume}{32}},
  \bibinfo{pages}{2302} (\bibinfo{year}{1985}).

\bibitem[{\citenamefont{An et~al.}(2008)\citenamefont{An, Subedi, and
  Singh}}]{ssc148-417}
\bibinfo{author}{\bibfnamefont{J.}~\bibnamefont{An}},
  \bibinfo{author}{\bibfnamefont{A.}~\bibnamefont{Subedi}}, \bibnamefont{and}
  \bibinfo{author}{\bibfnamefont{D.~J.} \bibnamefont{Singh}},
  \bibinfo{journal}{Solid State Commun.} \textbf{\bibinfo{volume}{148}},
  \bibinfo{pages}{417} (\bibinfo{year}{2008}).

\bibitem[{\citenamefont{Delaire et~al.}(2011)\citenamefont{Delaire, Ma, Marty,
  May, McGuire, Du, Singh, Podlesnyak, Ehlers, Lumsden et~al.}}]{nmat10-614}
\bibinfo{author}{\bibfnamefont{O.}~\bibnamefont{Delaire}},
  \bibinfo{author}{\bibfnamefont{J.}~\bibnamefont{Ma}},
  \bibinfo{author}{\bibfnamefont{K.}~\bibnamefont{Marty}},
  \bibinfo{author}{\bibfnamefont{A.~F.} \bibnamefont{May}},
  \bibinfo{author}{\bibfnamefont{M.~A.} \bibnamefont{McGuire}},
  \bibinfo{author}{\bibfnamefont{M.-H.} \bibnamefont{Du}},
  \bibinfo{author}{\bibfnamefont{D.~J.} \bibnamefont{Singh}},
  \bibinfo{author}{\bibfnamefont{A.}~\bibnamefont{Podlesnyak}},
  \bibinfo{author}{\bibfnamefont{G.}~\bibnamefont{Ehlers}},
  \bibinfo{author}{\bibfnamefont{M.~D.} \bibnamefont{Lumsden}},
  \bibnamefont{et~al.}, \bibinfo{journal}{Nat. Mater.}
  \textbf{\bibinfo{volume}{10}}, \bibinfo{pages}{614} (\bibinfo{year}{2011}).

\bibitem[{\citenamefont{Shiga et~al.}(2012)\citenamefont{Shiga, Shiomi, Ma,
  Delaire, Radzynski, Lusakowski, Esfarjani, and Chen}}]{prb85-155203}
\bibinfo{author}{\bibfnamefont{T.}~\bibnamefont{Shiga}},
  \bibinfo{author}{\bibfnamefont{J.}~\bibnamefont{Shiomi}},
  \bibinfo{author}{\bibfnamefont{J.}~\bibnamefont{Ma}},
  \bibinfo{author}{\bibfnamefont{O.}~\bibnamefont{Delaire}},
  \bibinfo{author}{\bibfnamefont{T.}~\bibnamefont{Radzynski}},
  \bibinfo{author}{\bibfnamefont{A.}~\bibnamefont{Lusakowski}},
  \bibinfo{author}{\bibfnamefont{K.}~\bibnamefont{Esfarjani}},
  \bibnamefont{and} \bibinfo{author}{\bibfnamefont{G.}~\bibnamefont{Chen}},
  \bibinfo{journal}{Phys. Rev. B} \textbf{\bibinfo{volume}{85}},
  \bibinfo{pages}{155203} (\bibinfo{year}{2012}).

\bibitem[{\citenamefont{Zhang et~al.}(2009)\citenamefont{Zhang, Ke, Chen, Yang,
  and Kent}}]{prb80-024304}
\bibinfo{author}{\bibfnamefont{Y.}~\bibnamefont{Zhang}},
  \bibinfo{author}{\bibfnamefont{X.}~\bibnamefont{Ke}},
  \bibinfo{author}{\bibfnamefont{C.}~\bibnamefont{Chen}},
  \bibinfo{author}{\bibfnamefont{J.}~\bibnamefont{Yang}}, \bibnamefont{and}
  \bibinfo{author}{\bibfnamefont{P.~R.~C.} \bibnamefont{Kent}},
  \bibinfo{journal}{Phys. Rev. B} \textbf{\bibinfo{volume}{80}},
  \bibinfo{pages}{024304} (\bibinfo{year}{2009}).

\bibitem[{\citenamefont{Romero et~al.}(2015)\citenamefont{Romero, Gross,
  Verstraete, and Hellman}}]{prb91-214310}
\bibinfo{author}{\bibfnamefont{A.~H.} \bibnamefont{Romero}},
  \bibinfo{author}{\bibfnamefont{E.~K.~U.} \bibnamefont{Gross}},
  \bibinfo{author}{\bibfnamefont{M.~J.} \bibnamefont{Verstraete}},
  \bibnamefont{and} \bibinfo{author}{\bibfnamefont{O.}~\bibnamefont{Hellman}},
  \bibinfo{journal}{Phys. Rev. B} \textbf{\bibinfo{volume}{91}},
  \bibinfo{pages}{214310} (\bibinfo{year}{2015}).

\bibitem[{\citenamefont{Skelton et~al.}(2014)\citenamefont{Skelton, Parker,
  Togo, Tanaka, and Walsh}}]{prb89-205203}
\bibinfo{author}{\bibfnamefont{J.~M.} \bibnamefont{Skelton}},
  \bibinfo{author}{\bibfnamefont{S.~C.} \bibnamefont{Parker}},
  \bibinfo{author}{\bibfnamefont{A.}~\bibnamefont{Togo}},
  \bibinfo{author}{\bibfnamefont{I.}~\bibnamefont{Tanaka}}, \bibnamefont{and}
  \bibinfo{author}{\bibfnamefont{A.}~\bibnamefont{Walsh}},
  \bibinfo{journal}{Phys. Rev. B} \textbf{\bibinfo{volume}{89}},
  \bibinfo{pages}{205203} (\bibinfo{year}{2014}).

\bibitem[{\citenamefont{Tian et~al.}(2012)\citenamefont{Tian, Garg, Esfarjani,
  Shiga, Shiomi, and Chen}}]{pbsete}
\bibinfo{author}{\bibfnamefont{Z.}~\bibnamefont{Tian}},
  \bibinfo{author}{\bibfnamefont{J.}~\bibnamefont{Garg}},
  \bibinfo{author}{\bibfnamefont{K.}~\bibnamefont{Esfarjani}},
  \bibinfo{author}{\bibfnamefont{T.}~\bibnamefont{Shiga}},
  \bibinfo{author}{\bibfnamefont{J.}~\bibnamefont{Shiomi}}, \bibnamefont{and}
  \bibinfo{author}{\bibfnamefont{G.}~\bibnamefont{Chen}},
  \bibinfo{journal}{Phys. Rev. B} \textbf{\bibinfo{volume}{85}},
  \bibinfo{pages}{184303} (\bibinfo{year}{2012}).

\bibitem[{\citenamefont{Hohnke et~al.}(1972)\citenamefont{Hohnke, Holloway, and
  Kaiser}}]{pbgete}
\bibinfo{author}{\bibfnamefont{D.~K.} \bibnamefont{Hohnke}},
  \bibinfo{author}{\bibfnamefont{H.}~\bibnamefont{Holloway}}, \bibnamefont{and}
  \bibinfo{author}{\bibfnamefont{S.}~\bibnamefont{Kaiser}},
  \bibinfo{journal}{J. Phys. Chem. Solids} \textbf{\bibinfo{volume}{33}},
  \bibinfo{pages}{2053} (\bibinfo{year}{1972}).

\bibitem[{\citenamefont{Akhtar et~al.}(2011)\citenamefont{Akhtar, Afzaal,
  Vincent, Burton, Hillier, and O{'}Brien}}]{pbs-polymer}
\bibinfo{author}{\bibfnamefont{J.}~\bibnamefont{Akhtar}},
  \bibinfo{author}{\bibfnamefont{M.}~\bibnamefont{Afzaal}},
  \bibinfo{author}{\bibfnamefont{M.~A.} \bibnamefont{Vincent}},
  \bibinfo{author}{\bibfnamefont{N.~A.} \bibnamefont{Burton}},
  \bibinfo{author}{\bibfnamefont{I.~H.} \bibnamefont{Hillier}},
  \bibnamefont{and}
  \bibinfo{author}{\bibfnamefont{P.}~\bibnamefont{O{'}Brien}},
  \bibinfo{journal}{Chem. Commun.} \textbf{\bibinfo{volume}{47}},
  \bibinfo{pages}{1991} (\bibinfo{year}{2011}).

\bibitem[{\citenamefont{Grozdanov}(1994)}]{pbse-pbs-polymer}
\bibinfo{author}{\bibfnamefont{I.}~\bibnamefont{Grozdanov}},
  \bibinfo{journal}{Chem. Lett.} \textbf{\bibinfo{volume}{23}},
  \bibinfo{pages}{551} (\bibinfo{year}{1994}).

\bibitem[{\citenamefont{Payne et~al.}(1992)\citenamefont{Payne, Teter, Allan,
  Arias, and Joannopoulos}}]{rmp64-1045}
\bibinfo{author}{\bibfnamefont{M.~C.} \bibnamefont{Payne}},
  \bibinfo{author}{\bibfnamefont{M.~P.} \bibnamefont{Teter}},
  \bibinfo{author}{\bibfnamefont{D.~C.} \bibnamefont{Allan}},
  \bibinfo{author}{\bibfnamefont{T.~A.} \bibnamefont{Arias}}, \bibnamefont{and}
  \bibinfo{author}{\bibfnamefont{J.~D.} \bibnamefont{Joannopoulos}},
  \bibinfo{journal}{Rev. Mod. Phys.} \textbf{\bibinfo{volume}{64}},
  \bibinfo{pages}{1045} (\bibinfo{year}{1992}).

\bibitem[{\citenamefont{Srivastava}(1990)}]{srivastava}
\bibinfo{author}{\bibfnamefont{G.~P.} \bibnamefont{Srivastava}},
  \emph{\bibinfo{title}{The Physics of Phonons}} (\bibinfo{publisher}{Taylor \&
  Francis Group, New York, U.S.A.}, \bibinfo{year}{1990}).

\bibitem[{\citenamefont{Broido et~al.}(2007)\citenamefont{Broido, Malorny,
  Birner, Mingo, and Stewart}}]{apl91-231922}
\bibinfo{author}{\bibfnamefont{D.~A.} \bibnamefont{Broido}},
  \bibinfo{author}{\bibfnamefont{M.}~\bibnamefont{Malorny}},
  \bibinfo{author}{\bibfnamefont{G.}~\bibnamefont{Birner}},
  \bibinfo{author}{\bibfnamefont{N.}~\bibnamefont{Mingo}}, \bibnamefont{and}
  \bibinfo{author}{\bibfnamefont{D.~A.} \bibnamefont{Stewart}},
  \bibinfo{journal}{Appl. Phys. Lett.} \textbf{\bibinfo{volume}{91}},
  \bibinfo{eid}{231922} (\bibinfo{year}{2007}).

\bibitem[{\citenamefont{Lindsay et~al.}(2013)\citenamefont{Lindsay, Broido, and
  Reinecke}}]{prb87-165201}
\bibinfo{author}{\bibfnamefont{L.}~\bibnamefont{Lindsay}},
  \bibinfo{author}{\bibfnamefont{D.~A.} \bibnamefont{Broido}},
  \bibnamefont{and} \bibinfo{author}{\bibfnamefont{T.~L.}
  \bibnamefont{Reinecke}}, \bibinfo{journal}{Phys. Rev. B}
  \textbf{\bibinfo{volume}{87}}, \bibinfo{pages}{165201}
  (\bibinfo{year}{2013}).

\bibitem[{\citenamefont{Togo et~al.}(2008)\citenamefont{Togo, Oba, and
  Tanaka}}]{phonopy}
\bibinfo{author}{\bibfnamefont{A.}~\bibnamefont{Togo}},
  \bibinfo{author}{\bibfnamefont{F.}~\bibnamefont{Oba}}, \bibnamefont{and}
  \bibinfo{author}{\bibfnamefont{I.}~\bibnamefont{Tanaka}},
  \bibinfo{journal}{Phys. Rev. B} \textbf{\bibinfo{volume}{78}},
  \bibinfo{pages}{134106} (\bibinfo{year}{2008}).

\bibitem[{\citenamefont{Chaput et~al.}(2011)\citenamefont{Chaput, Togo, Tanaka,
  and Hug}}]{prb84-094302}
\bibinfo{author}{\bibfnamefont{L.}~\bibnamefont{Chaput}},
  \bibinfo{author}{\bibfnamefont{A.}~\bibnamefont{Togo}},
  \bibinfo{author}{\bibfnamefont{I.}~\bibnamefont{Tanaka}}, \bibnamefont{and}
  \bibinfo{author}{\bibfnamefont{G.}~\bibnamefont{Hug}},
  \bibinfo{journal}{Phys. Rev. B} \textbf{\bibinfo{volume}{84}},
  \bibinfo{pages}{094302} (\bibinfo{year}{2011}).

\bibitem[{\citenamefont{Gonze et~al.}(2009)\citenamefont{Gonze, Amadon,
  Anglade, Beuken, Bottin, Boulanger, Bruneval, Caliste, Caracas, Cote
  et~al.}}]{abinit}
\bibinfo{author}{\bibfnamefont{X.}~\bibnamefont{Gonze}},
  \bibinfo{author}{\bibfnamefont{B.}~\bibnamefont{Amadon}},
  \bibinfo{author}{\bibfnamefont{P.-M.} \bibnamefont{Anglade}},
  \bibinfo{author}{\bibfnamefont{J.-M.} \bibnamefont{Beuken}},
  \bibinfo{author}{\bibfnamefont{F.}~\bibnamefont{Bottin}},
  \bibinfo{author}{\bibfnamefont{P.}~\bibnamefont{Boulanger}},
  \bibinfo{author}{\bibfnamefont{F.}~\bibnamefont{Bruneval}},
  \bibinfo{author}{\bibfnamefont{D.}~\bibnamefont{Caliste}},
  \bibinfo{author}{\bibfnamefont{R.}~\bibnamefont{Caracas}},
  \bibinfo{author}{\bibfnamefont{M.}~\bibnamefont{Cote}}, \bibnamefont{et~al.},
  \bibinfo{journal}{Comput. Phys. Commun.} \textbf{\bibinfo{volume}{180}},
  \bibinfo{pages}{2582 } (\bibinfo{year}{2009}).

\bibitem[{\citenamefont{Hartwigsen et~al.}(1998)\citenamefont{Hartwigsen,
  Goedecker, and Hutter}}]{hgh}
\bibinfo{author}{\bibfnamefont{C.}~\bibnamefont{Hartwigsen}},
  \bibinfo{author}{\bibfnamefont{S.}~\bibnamefont{Goedecker}},
  \bibnamefont{and} \bibinfo{author}{\bibfnamefont{J.}~\bibnamefont{Hutter}},
  \bibinfo{journal}{Phys. Rev. B} \textbf{\bibinfo{volume}{58}},
  \bibinfo{pages}{3641} (\bibinfo{year}{1998}).

\bibitem[{sup({\natexlab{a}})}]{suppl}
\bibinfo{howpublished}{See Supplemental Material at [url] for more details
  about our computational approach, which includes
  Refs.~\cite{natcomm,prb86-174307,nonlinear,dfpt-abinit}}.

\bibitem[{\citenamefont{Savi\'c et~al.}(2013)\citenamefont{Savi\'c, Donadio,
  Gygi, and Galli}}]{ivana-apl}
\bibinfo{author}{\bibfnamefont{I.}~\bibnamefont{Savi\'c}},
  \bibinfo{author}{\bibfnamefont{D.}~\bibnamefont{Donadio}},
  \bibinfo{author}{\bibfnamefont{F.}~\bibnamefont{Gygi}}, \bibnamefont{and}
  \bibinfo{author}{\bibfnamefont{G.}~\bibnamefont{Galli}},
  \bibinfo{journal}{Appl. Phys. Lett.} \textbf{\bibinfo{volume}{102}},
  \bibinfo{pages}{073113} (\bibinfo{year}{2013}).

\bibitem[{\citenamefont{He et~al.}(2012)\citenamefont{He, Savi\'c, Donadio, and
  Galli}}]{pccp}
\bibinfo{author}{\bibfnamefont{Y.}~\bibnamefont{He}},
  \bibinfo{author}{\bibfnamefont{I.}~\bibnamefont{Savi\'c}},
  \bibinfo{author}{\bibfnamefont{D.}~\bibnamefont{Donadio}}, \bibnamefont{and}
  \bibinfo{author}{\bibfnamefont{G.}~\bibnamefont{Galli}},
  \bibinfo{journal}{Phys. Chem. Chem. Phys.} \textbf{\bibinfo{volume}{14}},
  \bibinfo{pages}{16209} (\bibinfo{year}{2012}).

\bibitem[{sup({\natexlab{b}})}]{suppl2}
\bibinfo{howpublished}{See Supplemental Material at [url] for the figures
  illustrating some of the effects discussed in the paper, which includes
  Refs.~\cite{ins,pbte-exp,pbte-exp2,prb72-014308}}.

\bibitem[{\citenamefont{Murakami et~al.}(2013)\citenamefont{Murakami, Shiga,
  Hori, Esfarjani, and Shiomi}}]{pbsete-better-disorder}
\bibinfo{author}{\bibfnamefont{T.}~\bibnamefont{Murakami}},
  \bibinfo{author}{\bibfnamefont{T.}~\bibnamefont{Shiga}},
  \bibinfo{author}{\bibfnamefont{T.}~\bibnamefont{Hori}},
  \bibinfo{author}{\bibfnamefont{K.}~\bibnamefont{Esfarjani}},
  \bibnamefont{and} \bibinfo{author}{\bibfnamefont{J.}~\bibnamefont{Shiomi}},
  \bibinfo{journal}{Europhys. Lett.} \textbf{\bibinfo{volume}{102}},
  \bibinfo{pages}{46002} (\bibinfo{year}{2013}).

\bibitem[{\citenamefont{Klemens}(1955)}]{klemens-point-def}
\bibinfo{author}{\bibfnamefont{P.~G.} \bibnamefont{Klemens}},
  \bibinfo{journal}{Proc. Phys. Soc. A} \textbf{\bibinfo{volume}{68}},
  \bibinfo{pages}{1113} (\bibinfo{year}{1955}).

\bibitem[{\citenamefont{Gonze et~al.}(1994)\citenamefont{Gonze, Charlier,
  Allan, and Teter}}]{prb50-13035}
\bibinfo{author}{\bibfnamefont{X.}~\bibnamefont{Gonze}},
  \bibinfo{author}{\bibfnamefont{J.-C.} \bibnamefont{Charlier}},
  \bibinfo{author}{\bibfnamefont{D.~C.} \bibnamefont{Allan}}, \bibnamefont{and}
  \bibinfo{author}{\bibfnamefont{M.~P.} \bibnamefont{Teter}},
  \bibinfo{journal}{Phys. Rev. B} \textbf{\bibinfo{volume}{50}},
  \bibinfo{pages}{13035} (\bibinfo{year}{1994}).

\bibitem[{\citenamefont{Lindsay and Broido}(2008)}]{threephonon}
\bibinfo{author}{\bibfnamefont{L.}~\bibnamefont{Lindsay}} \bibnamefont{and}
  \bibinfo{author}{\bibfnamefont{D.~A.} \bibnamefont{Broido}},
  \bibinfo{journal}{J. Phys.: Condens. Matter} \textbf{\bibinfo{volume}{20}},
  \bibinfo{pages}{165209} (\bibinfo{year}{2008}).

\bibitem[{\citenamefont{Devyatkova and Smirnov}(1962)}]{pbte-exp}
\bibinfo{author}{\bibfnamefont{E.~D.} \bibnamefont{Devyatkova}}
  \bibnamefont{and} \bibinfo{author}{\bibfnamefont{I.~A.}
  \bibnamefont{Smirnov}}, \bibinfo{journal}{Sov. Phys. Solid State, USSR}
  \textbf{\bibinfo{volume}{3}}, \bibinfo{pages}{1666} (\bibinfo{year}{1962}).

\bibitem[{\citenamefont{Devyatkova and Tikhonov}(1965)}]{pbsete-exp}
\bibinfo{author}{\bibfnamefont{E.~D.} \bibnamefont{Devyatkova}}
  \bibnamefont{and} \bibinfo{author}{\bibfnamefont{V.~V.}
  \bibnamefont{Tikhonov}}, \bibinfo{journal}{Sov. Phys. Solid State, USSR}
  \textbf{\bibinfo{volume}{7}}, \bibinfo{pages}{1427} (\bibinfo{year}{1965}).

\bibitem[{kap({\natexlab{a}})}]{kappa-gete}
\bibinfo{howpublished}{Our calculated value for the isotropically averaged
  lattice thermal conductivity of GeTe at $450$~K is $3.27$~W/mK.}

\bibitem[{kap({\natexlab{b}})}]{kappa-vs-t-pbgete}
\bibinfo{howpublished}{The larger mass difference between Pb and Ge than that
  between Se and Te leads to the stronger alloy scattering in
  Pb$_{0.51}$Ge$_{0.49}$Te than in PbSe$_{0.5}$Te$_{0.5}$ alloys. Consequently,
  our calculated $\kappa$ of Pb$_{0.51}$Ge$_{0.49}$Te exhibits the $T^{-1/2}$
  rather than the $T^{-1}$ temperature dependence as
  PbSe$_{0.5}$Te$_{0.5}$~\cite{klemens-point-def2}, see Fig.~3~(a).}

\bibitem[{\citenamefont{Rabe and Joannopoulos}(1987)}]{PhysRevB.36.6631}
\bibinfo{author}{\bibfnamefont{K.~M.} \bibnamefont{Rabe}} \bibnamefont{and}
  \bibinfo{author}{\bibfnamefont{J.~D.} \bibnamefont{Joannopoulos}},
  \bibinfo{journal}{Phys. Rev. B} \textbf{\bibinfo{volume}{36}},
  \bibinfo{pages}{6631} (\bibinfo{year}{1987}).

\bibitem[{\citenamefont{Bo\v{z}in et~al.}(2010)\citenamefont{Bo\v{z}in,
  Malliakas, Souvatzis, Proffen, Spaldin, Kanatzidis, and Billinge}}]{bozin}
\bibinfo{author}{\bibfnamefont{E.~S.} \bibnamefont{Bo\v{z}in}},
  \bibinfo{author}{\bibfnamefont{C.~D.} \bibnamefont{Malliakas}},
  \bibinfo{author}{\bibfnamefont{P.}~\bibnamefont{Souvatzis}},
  \bibinfo{author}{\bibfnamefont{T.}~\bibnamefont{Proffen}},
  \bibinfo{author}{\bibfnamefont{N.~A.} \bibnamefont{Spaldin}},
  \bibinfo{author}{\bibfnamefont{M.~G.} \bibnamefont{Kanatzidis}},
  \bibnamefont{and} \bibinfo{author}{\bibfnamefont{S.~J.~L.}
  \bibnamefont{Billinge}}, \bibinfo{journal}{Science}
  \textbf{\bibinfo{volume}{330}}, \bibinfo{pages}{1660} (\bibinfo{year}{2010}).

\bibitem[{\citenamefont{Chen et~al.}(2014)\citenamefont{Chen, Ai, and
  Marianetti}}]{marianetti}
\bibinfo{author}{\bibfnamefont{Y.}~\bibnamefont{Chen}},
  \bibinfo{author}{\bibfnamefont{X.}~\bibnamefont{Ai}}, \bibnamefont{and}
  \bibinfo{author}{\bibfnamefont{C.~A.} \bibnamefont{Marianetti}},
  \bibinfo{journal}{Phys. Rev. Lett.} \textbf{\bibinfo{volume}{113}},
  \bibinfo{pages}{105501} (\bibinfo{year}{2014}).

\bibitem[{\citenamefont{Li et~al.}(2014)\citenamefont{Li, Hellman, Ma, May,
  Cao, Chen, Christianson, Ehlers, Singh, Sales
  et~al.}}]{PhysRevLett.112.175501}
\bibinfo{author}{\bibfnamefont{C.~W.} \bibnamefont{Li}},
  \bibinfo{author}{\bibfnamefont{O.}~\bibnamefont{Hellman}},
  \bibinfo{author}{\bibfnamefont{J.}~\bibnamefont{Ma}},
  \bibinfo{author}{\bibfnamefont{A.~F.} \bibnamefont{May}},
  \bibinfo{author}{\bibfnamefont{H.~B.} \bibnamefont{Cao}},
  \bibinfo{author}{\bibfnamefont{X.}~\bibnamefont{Chen}},
  \bibinfo{author}{\bibfnamefont{A.~D.} \bibnamefont{Christianson}},
  \bibinfo{author}{\bibfnamefont{G.}~\bibnamefont{Ehlers}},
  \bibinfo{author}{\bibfnamefont{D.~J.} \bibnamefont{Singh}},
  \bibinfo{author}{\bibfnamefont{B.~C.} \bibnamefont{Sales}},
  \bibnamefont{et~al.}, \bibinfo{journal}{Phys. Rev. Lett.}
  \textbf{\bibinfo{volume}{112}}, \bibinfo{pages}{175501}
  (\bibinfo{year}{2014}).

\bibitem[{\citenamefont{Lu et~al.}(2015)\citenamefont{Lu, Li, Wang, Li, Liu,
  and Ao}}]{jac621-345}
\bibinfo{author}{\bibfnamefont{Z.}~\bibnamefont{Lu}},
  \bibinfo{author}{\bibfnamefont{J.}~\bibnamefont{Li}},
  \bibinfo{author}{\bibfnamefont{C.}~\bibnamefont{Wang}},
  \bibinfo{author}{\bibfnamefont{Y.}~\bibnamefont{Li}},
  \bibinfo{author}{\bibfnamefont{F.}~\bibnamefont{Liu}}, \bibnamefont{and}
  \bibinfo{author}{\bibfnamefont{W.}~\bibnamefont{Ao}}, \bibinfo{journal}{J.
  Alloy Compd.} \textbf{\bibinfo{volume}{621}}, \bibinfo{pages}{345 }
  (\bibinfo{year}{2015}).

\bibitem[{\citenamefont{Wu et~al.}(2014)\citenamefont{Wu, Zhao, Hao, Jiang,
  Zheng, Doak, Wu, Chi, Gelbstein, Uher et~al.}}]{pbgete-jacs}
\bibinfo{author}{\bibfnamefont{D.}~\bibnamefont{Wu}},
  \bibinfo{author}{\bibfnamefont{L.-D.} \bibnamefont{Zhao}},
  \bibinfo{author}{\bibfnamefont{S.}~\bibnamefont{Hao}},
  \bibinfo{author}{\bibfnamefont{Q.}~\bibnamefont{Jiang}},
  \bibinfo{author}{\bibfnamefont{F.}~\bibnamefont{Zheng}},
  \bibinfo{author}{\bibfnamefont{J.~W.} \bibnamefont{Doak}},
  \bibinfo{author}{\bibfnamefont{H.}~\bibnamefont{Wu}},
  \bibinfo{author}{\bibfnamefont{H.}~\bibnamefont{Chi}},
  \bibinfo{author}{\bibfnamefont{Y.}~\bibnamefont{Gelbstein}},
  \bibinfo{author}{\bibfnamefont{C.}~\bibnamefont{Uher}}, \bibnamefont{et~al.},
  \bibinfo{journal}{J. Am. Chem. Soc.} \textbf{\bibinfo{volume}{136}},
  \bibinfo{pages}{11412} (\bibinfo{year}{2014}).

\bibitem[{\citenamefont{Gelbstein et~al.}(2013)\citenamefont{Gelbstein,
  Davidow, Girard, Chung, and Kanatzidis}}]{gelbstein-adv-energy-mat}
\bibinfo{author}{\bibfnamefont{Y.}~\bibnamefont{Gelbstein}},
  \bibinfo{author}{\bibfnamefont{J.}~\bibnamefont{Davidow}},
  \bibinfo{author}{\bibfnamefont{S.~N.} \bibnamefont{Girard}},
  \bibinfo{author}{\bibfnamefont{D.~Y.} \bibnamefont{Chung}}, \bibnamefont{and}
  \bibinfo{author}{\bibfnamefont{M.}~\bibnamefont{Kanatzidis}},
  \bibinfo{journal}{Adv. Energy Mat.} \textbf{\bibinfo{volume}{3}},
  \bibinfo{pages}{815} (\bibinfo{year}{2013}).

\bibitem[{\citenamefont{Li et~al.}(2011)\citenamefont{Li, Li, Wang, Wang, Liu,
  and Ao}}]{solidstatesci}
\bibinfo{author}{\bibfnamefont{S.}~\bibnamefont{Li}},
  \bibinfo{author}{\bibfnamefont{J.}~\bibnamefont{Li}},
  \bibinfo{author}{\bibfnamefont{Q.}~\bibnamefont{Wang}},
  \bibinfo{author}{\bibfnamefont{L.}~\bibnamefont{Wang}},
  \bibinfo{author}{\bibfnamefont{F.}~\bibnamefont{Liu}}, \bibnamefont{and}
  \bibinfo{author}{\bibfnamefont{W.}~\bibnamefont{Ao}}, \bibinfo{journal}{Solid
  State Sci.} \textbf{\bibinfo{volume}{13}}, \bibinfo{pages}{399 }
  (\bibinfo{year}{2011}).

\bibitem[{\citenamefont{Schl\"uter et~al.}(1975)\citenamefont{Schl\"uter,
  Martinez, and Cohen}}]{PhysRevB.12.650}
\bibinfo{author}{\bibfnamefont{M.}~\bibnamefont{Schl\"uter}},
  \bibinfo{author}{\bibfnamefont{G.}~\bibnamefont{Martinez}}, \bibnamefont{and}
  \bibinfo{author}{\bibfnamefont{M.~L.} \bibnamefont{Cohen}},
  \bibinfo{journal}{Phys. Rev. B} \textbf{\bibinfo{volume}{12}},
  \bibinfo{pages}{650} (\bibinfo{year}{1975}).

\bibitem[{\citenamefont{Gonze and Lee}(1997)}]{dfpt-abinit}
\bibinfo{author}{\bibfnamefont{X.}~\bibnamefont{Gonze}} \bibnamefont{and}
  \bibinfo{author}{\bibfnamefont{C.}~\bibnamefont{Lee}},
  \bibinfo{journal}{Phys. Rev. B} \textbf{\bibinfo{volume}{55}},
  \bibinfo{pages}{10355} (\bibinfo{year}{1997}).

\bibitem[{\citenamefont{Pei et~al.}(2011)\citenamefont{Pei, Shi, LaLonde, Wang,
  Chen, and Snyder}}]{pbsete-nature}
\bibinfo{author}{\bibfnamefont{Y.}~\bibnamefont{Pei}},
  \bibinfo{author}{\bibfnamefont{X.}~\bibnamefont{Shi}},
  \bibinfo{author}{\bibfnamefont{A.}~\bibnamefont{LaLonde}},
  \bibinfo{author}{\bibfnamefont{H.}~\bibnamefont{Wang}},
  \bibinfo{author}{\bibfnamefont{L.}~\bibnamefont{Chen}}, \bibnamefont{and}
  \bibinfo{author}{\bibfnamefont{G.~J.} \bibnamefont{Snyder}},
  \bibinfo{journal}{Nature} \textbf{\bibinfo{volume}{473}}, \bibinfo{pages}{66}
  (\bibinfo{year}{2011}).

\bibitem[{\citenamefont{Heremans et~al.}(2008)\citenamefont{Heremans, Jovovic,
  Toberer, Saramat, Kurosaki, Charoenphakdee, Yamanaka, and Snyder}}]{resimp}
\bibinfo{author}{\bibfnamefont{J.~P.} \bibnamefont{Heremans}},
  \bibinfo{author}{\bibfnamefont{V.}~\bibnamefont{Jovovic}},
  \bibinfo{author}{\bibfnamefont{E.~S.} \bibnamefont{Toberer}},
  \bibinfo{author}{\bibfnamefont{A.}~\bibnamefont{Saramat}},
  \bibinfo{author}{\bibfnamefont{K.}~\bibnamefont{Kurosaki}},
  \bibinfo{author}{\bibfnamefont{A.}~\bibnamefont{Charoenphakdee}},
  \bibinfo{author}{\bibfnamefont{S.}~\bibnamefont{Yamanaka}}, \bibnamefont{and}
  \bibinfo{author}{\bibfnamefont{G.~J.} \bibnamefont{Snyder}},
  \bibinfo{journal}{Science} \textbf{\bibinfo{volume}{321}},
  \bibinfo{pages}{554} (\bibinfo{year}{2008}).

\bibitem[{\citenamefont{Li et~al.}(2015)\citenamefont{Li, Hong, May, Bansal,
  Chi, Hong, Ehlers, and Delaire}}]{delaire-snse}
\bibinfo{author}{\bibfnamefont{C.~W.} \bibnamefont{Li}},
  \bibinfo{author}{\bibfnamefont{J.}~\bibnamefont{Hong}},
  \bibinfo{author}{\bibfnamefont{A.~F.} \bibnamefont{May}},
  \bibinfo{author}{\bibfnamefont{D.}~\bibnamefont{Bansal}},
  \bibinfo{author}{\bibfnamefont{S.}~\bibnamefont{Chi}},
  \bibinfo{author}{\bibfnamefont{T.}~\bibnamefont{Hong}},
  \bibinfo{author}{\bibfnamefont{G.}~\bibnamefont{Ehlers}}, \bibnamefont{and}
  \bibinfo{author}{\bibfnamefont{O.}~\bibnamefont{Delaire}},
  \bibinfo{journal}{Nat. Phys.} \textbf{\bibinfo{volume}{11}},
  \bibinfo{pages}{1063} (\bibinfo{year}{2015}).

\bibitem[{\citenamefont{Zhao et~al.}(2014{\natexlab{b}})\citenamefont{Zhao, Lo,
  Zhang, Sun, Tan, Uher, Wolverton, Dravid, and Kanatzidis}}]{nature-snse}
\bibinfo{author}{\bibfnamefont{L.-D.} \bibnamefont{Zhao}},
  \bibinfo{author}{\bibfnamefont{S.-H.} \bibnamefont{Lo}},
  \bibinfo{author}{\bibfnamefont{Y.}~\bibnamefont{Zhang}},
  \bibinfo{author}{\bibfnamefont{H.}~\bibnamefont{Sun}},
  \bibinfo{author}{\bibfnamefont{G.}~\bibnamefont{Tan}},
  \bibinfo{author}{\bibfnamefont{C.}~\bibnamefont{Uher}},
  \bibinfo{author}{\bibfnamefont{C.}~\bibnamefont{Wolverton}},
  \bibinfo{author}{\bibfnamefont{V.~P.} \bibnamefont{Dravid}},
  \bibnamefont{and} \bibinfo{author}{\bibfnamefont{M.~G.}
  \bibnamefont{Kanatzidis}}, \bibinfo{journal}{Nature}
  \textbf{\bibinfo{volume}{508}}, \bibinfo{pages}{373}
  (\bibinfo{year}{2014}{\natexlab{b}}).

\bibitem[{\citenamefont{Lee et~al.}(2014)\citenamefont{Lee, Esfarjani, Luo,
  Zhou, Tian, and Chen}}]{natcomm}
\bibinfo{author}{\bibfnamefont{S.}~\bibnamefont{Lee}},
  \bibinfo{author}{\bibfnamefont{K.}~\bibnamefont{Esfarjani}},
  \bibinfo{author}{\bibfnamefont{T.}~\bibnamefont{Luo}},
  \bibinfo{author}{\bibfnamefont{J.}~\bibnamefont{Zhou}},
  \bibinfo{author}{\bibfnamefont{Z.}~\bibnamefont{Tian}}, \bibnamefont{and}
  \bibinfo{author}{\bibfnamefont{G.}~\bibnamefont{Chen}},
  \bibinfo{journal}{Nat. Comm.} \textbf{\bibinfo{volume}{5}},
  \bibinfo{pages}{3525} (\bibinfo{year}{2014}).

\bibitem[{\citenamefont{Li et~al.}(2012)\citenamefont{Li, Lindsay, Broido,
  Stewart, and Mingo}}]{prb86-174307}
\bibinfo{author}{\bibfnamefont{W.}~\bibnamefont{Li}},
  \bibinfo{author}{\bibfnamefont{L.}~\bibnamefont{Lindsay}},
  \bibinfo{author}{\bibfnamefont{D.~A.} \bibnamefont{Broido}},
  \bibinfo{author}{\bibfnamefont{D.~A.} \bibnamefont{Stewart}},
  \bibnamefont{and} \bibinfo{author}{\bibfnamefont{N.}~\bibnamefont{Mingo}},
  \bibinfo{journal}{Phys. Rev. B} \textbf{\bibinfo{volume}{86}},
  \bibinfo{pages}{174307} (\bibinfo{year}{2012}).

\bibitem[{\citenamefont{Wang et~al.}(2010)\citenamefont{Wang, Wang, Wang, Mei,
  Shang, Chen, and Liu}}]{nonlinear}
\bibinfo{author}{\bibfnamefont{Y.}~\bibnamefont{Wang}},
  \bibinfo{author}{\bibfnamefont{J.~J.} \bibnamefont{Wang}},
  \bibinfo{author}{\bibfnamefont{W.~Y.} \bibnamefont{Wang}},
  \bibinfo{author}{\bibfnamefont{Z.~G.} \bibnamefont{Mei}},
  \bibinfo{author}{\bibfnamefont{S.~L.} \bibnamefont{Shang}},
  \bibinfo{author}{\bibfnamefont{L.~Q.} \bibnamefont{Chen}}, \bibnamefont{and}
  \bibinfo{author}{\bibfnamefont{Z.~K.} \bibnamefont{Liu}},
  \bibinfo{journal}{J. Phys.: Condens. Matter} \textbf{\bibinfo{volume}{22}},
  \bibinfo{pages}{202201} (\bibinfo{year}{2010}).

\bibitem[{\citenamefont{Cochran et~al.}(1966)\citenamefont{Cochran, Crowley,
  Dolling, and Elcombe}}]{ins}
\bibinfo{author}{\bibfnamefont{W.}~\bibnamefont{Cochran}},
  \bibinfo{author}{\bibfnamefont{R.~A.} \bibnamefont{Crowley}},
  \bibinfo{author}{\bibfnamefont{G.}~\bibnamefont{Dolling}}, \bibnamefont{and}
  \bibinfo{author}{\bibfnamefont{M.~M.} \bibnamefont{Elcombe}},
  \bibinfo{journal}{Proc. R. Soc. Lond. A} \textbf{\bibinfo{volume}{293}},
  \bibinfo{pages}{433} (\bibinfo{year}{1966}).

\bibitem[{\citenamefont{El-Sharkawy et~al.}(1983)\citenamefont{El-Sharkawy,
  El-Azm, Kenawy, Hillal, and Abu-Basha}}]{pbte-exp2}
\bibinfo{author}{\bibfnamefont{A.~A.} \bibnamefont{El-Sharkawy}},
  \bibinfo{author}{\bibfnamefont{A.~M.~A.} \bibnamefont{El-Azm}},
  \bibinfo{author}{\bibfnamefont{M.~I.} \bibnamefont{Kenawy}},
  \bibinfo{author}{\bibfnamefont{A.~S.} \bibnamefont{Hillal}},
  \bibnamefont{and} \bibinfo{author}{\bibfnamefont{H.~M.}
  \bibnamefont{Abu-Basha}}, \bibinfo{journal}{Int. J. Thermophys.}
  \textbf{\bibinfo{volume}{4}}, \bibinfo{pages}{261} (\bibinfo{year}{1983}).

\bibitem[{\citenamefont{Broido et~al.}(2005)\citenamefont{Broido, Ward, and
  Mingo}}]{prb72-014308}
\bibinfo{author}{\bibfnamefont{D.~A.} \bibnamefont{Broido}},
  \bibinfo{author}{\bibfnamefont{A.}~\bibnamefont{Ward}}, \bibnamefont{and}
  \bibinfo{author}{\bibfnamefont{N.}~\bibnamefont{Mingo}},
  \bibinfo{journal}{Phys. Rev. B} \textbf{\bibinfo{volume}{72}},
  \bibinfo{pages}{014308} (\bibinfo{year}{2005}).

\bibitem[{\citenamefont{Klemens}(1960)}]{klemens-point-def2}
\bibinfo{author}{\bibfnamefont{P.~G.} \bibnamefont{Klemens}},
  \bibinfo{journal}{Phys. Rev.} \textbf{\bibinfo{volume}{119}},
  \bibinfo{pages}{507} (\bibinfo{year}{1960}).

\end{thebibliography}

\end{document}